\documentclass[11pt]{article}
\input{epsf.sty}
\usepackage{hyperref}
\usepackage{amsmath, amssymb} 
\usepackage[pdftex]{graphicx}
\newtheorem {thm}{Theorem}[section]
\newtheorem {prop}[thm]{Proposition}
\newtheorem {lem}[thm]{Lemma}
\newtheorem {cor}[thm]{Corollary}
\newtheorem {defn}[thm]{Definition}

\def\Cox{\hfill \Box}

\def\R{{\Bbb R}}
\def\P{{\Bbb P}}

\def\0{{\bf 0}}

\def\sb{{\subset}}

\def\a{\alpha}

\def\b{\beta}

\def\d{\delta}

\def\e{\varepsilon}

\def\phi{\varphi}

\def\g{\gamma}

\def\k{\kappa}

\def\r{\rho}

\def\s{\sigma}

\def\t{\tau}

\def\x{\xi}

\def\o{\omega}

\def\L{\Lambda}

\def\G{\Gamma}

\def\O{\Omega}

\def\T{\T}

\def\HH{{\cal H}}
\def\PP{{\cal P}}
\def\NN{{\cal N}}

\def\GG{{\cal G}}

\begin{document}

\title{Metastates in finite-type mean-field models:\\
visibility, invisibility,  \\ 
and random restoration of symmetry} 

\author{
Giulio Iacobelli\thanks{
Department of Mathematics and Computing Sciences, University of Groningen, 
Nijenborgh 9,   
9747 AC Groningen, 
The Netherlands, 
\texttt{G.Iacobelli@rug.nl},
}\,\,
   and
Christof K\"ulske
%
\thanks{Ruhr-Universit\"at Bochum, Fakult\"at f\"ur Mathematik, Universit\"atsstra\ss e
150, 44780 Bochum, Germany, \texttt{Christof.Kuelske@rub.de, }
\texttt{http://www.ruhr-uni-bochum.de/ffm/Lehrstuehle/Kuelske/kuelske.html } 
 %
} 
}

\maketitle

\begin{abstract} 
We consider a general class of disordered mean-field models where 
both the spin variables and disorder variables $\eta$ take finitely many values. 
To investigate the size-dependence in the phase-transition regime 
we construct the metastate describing the probabilities to 
find a large system close to a particular convex combination of the pure 
infinite-volume states. 
We show that, under a non-degeneracy assumption, only pure states $j$ are seen, with 
non-random probability weights $w_j$ for which we derive explicit expressions in terms 
of interactions and distributions of the disorder variables. We provide a geometric 
construction distinguishing invisible states (having $w_j=0$) from visible ones. 
As a further consequence we show that,  
in the case where precisely two pure states are 
available, these must necessarily occur with the same weight, even if the model has 
no obvious symmetry relating the two. 
\end{abstract}

\smallskip
\noindent {\bf AMS 2000 subject classification:}  82B44, 82B26, 60K35.
\bigskip 

{\em Keywords:} 
Gibbs measures, mean-field systems, disordered systems, metastates, Ising model, Potts model.

\section{Introduction} \label{sect:intro}

Dealing with phase transitions in the theory of Gibbs measures of disordered systems 
is usually not an easy task. First of all, one likes to understand which are the possible 
phases and how do they depend on the realization of the disorder. This can be a formidable 
task, even for mean field models, as the history of the SK-model shows. 
Secondly, even if we suppose that the phases are identified, it is not a priori clear what 
role they will play for the typical behavior of a large but finite system. 
Indeed, in a regime where there are competing extremal phases (say a plus and a minus phase 
in a random field model) it may depend on the realization of the disorder variables 
which of the convex combinations the system will be close to in equilibrium. Some of the possible 
infinite-volume equilibrium states might not even show up in a typical large volume.  
To make sense of these questions, the concept of a metastate has been 
invented by Aizenman, Wehr \cite{AiWe90}, Newman and Stein \cite{NeSt97,Ne97},  being a probability measure 
which gives the weights in the large volume asymptotics to find a system close 
to one of the possible candidates among the Gibbs measures. We stress that the metastate
is an notion describing purely the equilibrium behavior. 
Yet another type of questions is of dynamic nature: Suppose a system 
undergoes a Glauber dynamics, how much time will it need to go from an initial 
state to it global free energy minimum? For some initial states this time is exceptionally 
long, a phenomenon called metastability (not to be confused with the notion of a metastate), 
and again, to derive precise asymptotic saying how long, 
examples of mean field systems have been very instructive model systems \cite{DaHo96,BiBoIo09}. 

On the lattice the metastate has been shown to be a useful concept in spinglasses 
by Newman and Stein \cite{NeSt01} and Arguin, Damron, Newman, Stein 
\cite{ArDaNeSt09} who showed that there is only one groundstate pair in the two-dimensional 
Edwards-Anderson model in the half-plane, 
using translation-ergodicity and Burton-Keane type of arguments. 

Explicit constructions for lattice models are difficult (see however \cite{EnNeSc06} where
the influence of random boundary conditions on an Ising model 
was analysed), but possible in mean-field models. Previously treated examples are given 
in very specific models, namely the symmetric random field Ising model and Hopfield
model with finite or a growing 
number of patterns \cite{Ku97,Ku98,Ku98b,BoGa98,BoEnNi99}. 

In this paper we aim for completeness in a particular direction, namely disordered  
mean field models with finitely many values for both spin and disorder variables. 
Such models include in particular the random-field Curie-Weiss Ising model and 
Potts-type Curie-Weiss random field models with or without symmetries 
in Hamiltonians or random field distributions. 
What we aim for is the abstract construction of the phase diagram, embellished with 
probability weights giving us the appearance of the candidate states.  
That is,  we first say {\em which} states are available. This, for disordered mean-field models comes from 
an investigation of the corresponding 
free energy (resp. rate functions) and is a standard thing. 
Next and new in our paper is the additional information on the {\em weights} with which they occur, 
and the proof of the validity of a corresponding approximate extreme decomposition, 
asymptotically for large volumes. This is then is then cast in the metastate formulation.  
The weights are obtained by studying the distribution of the free energy fluctuations 
w.r.t. to the disorder variables entering. 
Will the same type of results be true for corresponding lattice models 
at low temperatures at phase coexistence? We believe yes, but a proof will 
have to build around sophisticated expansion techniques and be technically 
rather challenging.  One would need to  show first the coexistence of 
states (as it was done for the random field Ising model 
in \cite{BrKu88}), and then the dominance of one of the available states over the 
others for typical realizations of the disorder. The mean field results should 
provide guidance for that, and moreover 
we believe that they are a rather nice complete example for a limit theorem 
in statistical mechanics.    
\bigskip

\subsection{The models: Mean-Field models with local disorder }

These are the models we consider. At each site $i=1,\dots, n$ there is a spin variable 
$\s(i)$ taking values in a finite set $E$ and a disorder   variable $\eta(i)$ taking values 
in the finite (possibly different) set $E'$.  We write $\PP(E)$ for the set of probability measures 
on $E$, and use similar notation for other spaces.  
 We write $L_n=\frac{1}{n}\sum_{i=1}^n \d_{\s(i)}\in \PP(E)$ for the (total) empirical measure 
of the spins and consider a twice continuously differentiable function $F$ on $\PP(E)$. The influence of the disorder 
variables on the Gibbs measures for the spins is through the local a priori measures $\a[b]\in \PP(E)$, 
for any possible type of the disorder $b\in E'$.  Hence the present analysis excludes models with 
disorder entering the interaction such as e.g. the Hopfield model treated in \cite{BoGa98,Ku97}.

\begin{defn}\label{132}
The mean-field model with Hamiltonian $n F(\nu)$ and a priori measures 
$\a[b]\in \PP(E)$, for all $b\in E'$,  is given by the disorder-dependent finite-volume Gibbs measures 
\begin{equation}
\begin{split}\label{1.3.2}
&\mu_{F,n}[\eta(1),\dots,\eta(n)](\s(1)=\o(1),\dots,\s(n)=\o(n))\cr
&=\frac{1}{Z_{F,n}[\eta(1),\dots,\eta(n)]}\exp\left(- n F\left(L^\o_n\right) \right)\prod_{i=1}^n\a[\eta_i](\o_i)\cr 
\end{split} 
\end{equation}
together with the prescription of a probability distribution $\pi\in \PP(E')$ for the disorder 
variables according to which they are chosen independently over the sites. We assume $\pi(b)>0$ 
for all $b\in E'$. 
\end{defn}

To summarize, our model depends on the triple of parameters $(F, \a, \pi)$ of:  
mean-field interaction $F$, a priori measures $\a=(\a[b])_{b\in E'}$, and disorder distribution $\pi$.

We need to introduce more notations. 
Given $\eta$, we write  
$$\L_n(b)=\{ i \in\{ 1,2,\dots, n\}; \eta(i)=b\}$$ for all $b\in E'$, for the $b$-like sites. 
 Write $$\hat \pi_n(b)=\frac{|\L_n(b)|}{n}$$ for the frequency of the $b$-like sites (empirical 
distribution of random field types.) 
Write $$\hat L_n(b)=\frac{1}{|\L_n(b)|}\sum_{i\in \L_n(b)}\d_{\s(i)}$$ for the empirical 
spin-distribution on the $b$-like sites. 
Write $\hat L_n=(\hat L_n(b))_{b\in E'}$ for the vector of empirical distributions. 
 The total empirical distribution is then the scalar product of $\hat\pi_n$ with the vector of empirical 
spin distributions 
\begin{equation}
\begin{split}\label{1.3.2}
L_n=\sum_{b\in E'}\hat \pi_n(b)\hat L_n(b)
\end{split} 
\end{equation}

\subsection{The metastate on the level of the states}

Let us jump into the following definition of a metastate, obtained 
by a conditioning procedure, which was given first by Aizenman and Wehr \cite{AiWe90}. 
There are different constructions of a metastate, but the present one will be the only 
one considered in the paper. 
This Aizenman-Wehr construction was related to a different and more intuitive 
construction as empirical averages of Gibbs measures 
along volume-(sub-)sequences by Newman and Stein. We refer to the monographs 
\cite{Ne97,Bo06}.  

\begin{defn} Assume that, for every bounded continuous $\Xi:\PP(E^\infty)\times (E')^\infty\rightarrow \R$  
the limit 
\begin{equation}
\begin{split}\label{nondeg3}
&\lim_{n\uparrow \infty}\int\P(d\eta)\Xi(\mu_{n}[\eta],\eta)=\int J(d\mu,d\eta)\Xi(\mu,\eta)\cr
\end{split} 
\end{equation}
exists. Then the conditional distribution $\k[\eta](d\mu):=  J(d\mu|\eta)$ 
is called the {\em AW-metastate on the level of the states}. 
\end{defn}
As it is common, continuity is meant in the following sense: 
A function on an infinite product of a finite space 
is continuous (w.r.t. local topology) if it is a uniform limit of local functions. 
For probability measures on $\PP(E^\infty)$ we use the weak topology (according 
to which a sequence of measures converges iff it converges on continuous test-functions), 
and for $\PP(E^\infty)\times (E')^\infty$, we use the product topology.

\subsection{Main Theorem} 

How do we get the possible equilibrium states of the system? 
They are obtained as solutions to the following minimization problem.

\begin{defn} Consider the free energy minimization problem 
\begin{equation}
\begin{split}\label{1.3.2}
&\hat \nu\mapsto \Phi[\pi](\hat \nu)
\end{split} 
\end{equation}
on $\PP(E)^{E'}$,  
with the free energy functional 
\begin{equation}
\begin{split}\label{fi}
&\Phi:\PP(E')\times \PP(E)^{E'}\mapsto \R\cr
&\Phi[\hat \pi](\hat \nu)= F\left(\sum_{b\in E'}\hat \pi(b)\hat \nu(b)
\right)+ \sum_{b}\hat \pi(b)S(\hat \nu(b)|\a[b])\cr 
\end{split} 
\end{equation}
where $S(p_1|p_2)=\sum_{a\in E}p_1(a)\log\frac{p_1(a)}{p_2(a)}$ is the relative entropy. 
We say that the random mean-field system obeys the {\em non-degeneracy condition 1)} 
if $\hat \nu \mapsto \Phi[\pi](\hat \nu)$ has a finite set of minimizers $M^*=M^*(F,\a,\pi)$ where 
all the eigenvalues of the Hessian are strictly positive. 
\end{defn}

It is very hard for a system not to satisfy this condition and we will assume 
in the following that it is satisfied. If it is true the vector of the empirical 
spin distributions of the system,  $\hat L_n$,  will concentrate around the set $M^*$. More than that, 
it may even concentrate on a smaller set.
The following 
theorem about the metastate will tell us how this concentration will take place and get the weights $w_j$.

Let $\hat \nu_j$ be a fixed element in $M^*$. 
Let us consider the linearization of the free energy functional at the fixed minimizers as a function of $\pi$, which reads 
\begin{equation}
\begin{split}\label{li-fi}
&\Phi[\tilde\pi](\hat\nu_j)-\Phi[\pi](\hat\nu_j)=-B_{j}[\tilde \pi -\pi]+ o(\Vert \tilde \pi-\pi\Vert ) \cr 
\end{split} 
\end{equation}
where 
\begin{equation}
\begin{split}\label{1.3.3}
&B_{j}[\tilde \pi -\pi]=-\left( d F_{\pi\cdot \hat \nu_j}\left(\sum_{b}(\tilde \pi(b)-\pi(b)){\hat \nu_j}(b)
\right)+\sum_{b}[\tilde \pi(b)-\pi(b)]S(\hat \nu_j(b)|\a[b])\right) \cr 
\end{split} 
\end{equation}
This defines an affine function on the 
tangent space of field type measures $T \PP(E')$ (i.e. vectors which sum up to zero), for any $j$.

\begin{defn} 
We call $B_{j}$ the stability vector of $\hat \nu_j$.  
We call 
\begin{equation}
\begin{split}\label{RJ}
&R_j:=\{x\in T\PP(E'), \langle x,  B_j \rangle>\max_{k\neq j}    \langle x, B_k \rangle \}
\end{split} 
\end{equation}
the stability region of $\hat \nu_j$. 
\end{defn}

Now comes our second condition. 

\begin{defn} We say the vector $B=(B_1,\dots,B_k)$ satisfies the 
{\em non-degeneracy condition 2)} if no different minimizers $j,j'$ have the same $B_j=B_{j'}$
\end{defn}

In other words the randomness lifts all symmetries. 
Note that this implies that $\Bigl(\bigcup_{j=1,\dots,k}R_j\Bigr)^c$ has zero Lebesgue measure 
in $T\PP(E')$. Indeed, if the map $j\mapsto  \langle x, B_j \rangle $ has no unique maximizer for 
fixed $x$,  then,  for some pair $j\neq k$ we have that $\langle x, B_j -B_k\rangle=0$. For fixed $j,k$ 
this set of $x$'s is a hyperplane (hence a measure zero set) since, by assumption,  $ B_j \neq B_k$.

We note the following simple but important geometric lemma.  
  
\begin{lem}  $R_j\neq \emptyset \Leftrightarrow B_j \in \text{ex}( \HH_{\text{conv}}\{B_1,\dots,B_k \} )$. 
 \end{lem}  
  
Here, for a subset $A \subset \R^d$,  $ \HH_{\text{conv}}(A) $  denotes the {\em convex hull} of $A$, 
that is the smallest convex set which contains $A$.  $\text{ex} (C)$, for a convex set $C$ denotes 
the extremal points of $C$,  that i those points which can not be written as a non-trivial convex 
combinations with points from $C$. In our case $\HH_{\text{conv}}\{B_1,\dots,B_k \}$ is 
a convex polyhedron and $\text{ex}( \HH_{\text{conv}}\{B_1,\dots,B_k \} )$ is the smallest 
set of points which generates it. \bigskip\bigskip

{\bf Proof of the Lemma.} We prove the implication $"\Rightarrow"$ by contradiction. 
Suppose that $B_j$ is not an extremal point.
Then it can be written as a non-trivial convex combination 
$B_j=\sum_{i}\a_i B_i$ with $\sum_{i=1}^k \a_i=1$, 
where $\a_i\geq 0$  and non-zero only for $B_i \in \text{ex}( \HH_{\text{conv}}\{B_1,\dots,B_k \} )$. 
Any vector  $x\in R_j$ satisfies $ \langle x,B_j \rangle >\langle x,B_i \rangle$ for all $i\neq j$ 
and hence  $   \langle x,B_j\rangle      =            \sum_{i}\a_i \langle x,B_j\rangle >   \sum_{i=1}^k   \a_i 
\langle x,B_i \rangle = \langle x,B_j \rangle $. This is a contradiction and hence $R_j=\emptyset$. 

To prove the opposite implication $"\Leftarrow"$ let us consider an extremal point $B_j$ 
and note the following: 
If $B_j \not\in \HH_{\text{conv}}\{B_1,\dots,B_{j-1},B_{j+1},\dots, B_k \}$ then, after 
a suitable translation and rotation, we can find 
coordinates such that the vectors take the form 
$B_j= (0 ,  \dots , 0 ,B_{j,d})$ and $B_i=(B'_i,B_{i,d})$ with $B_{j,d}> 0$ and 
$B_{i,d}\leq 0$ for $i\neq j$. (The latter statement follows from the fact that there is a 
separating hyperplane between $ \HH_{\text{conv}}\{B_1,\dots,B_{j-1},B_{j+1},\dots, B_k \} $ 
and the point $B_j$. This finite-dimensional version of the Hahn-Banach theorem 
is a classical result in geometry, see Theorem 1.2.4 in \cite{Ma02}. 
Having this separating hyperplane we choose the 
origin as the orthogonal projection of $B_{j}$ to this plane, the first coordinates as orthogonal coordinates 
inside the plane, and the last coordinate axis pointing in the direction of $B_j$.) 
The proof relies on the last two inequalities. 
Indeed we have, with the general notation $x=(x',x_d)\in \R^{d-1}\times \R$ that  
\begin{equation}
\begin{split}\label{nondeg3}
&R_j=\{x \in \R^d: \forall i\neq j \text{ holds } \langle x, B_j -B_i \rangle>0  \}\cr
&=\{x \in \R^d: \forall i\neq j \text{ holds } 
\langle x', B'_j -B'_i \rangle   + x_d(B_{j,d}-B_{i,d})>    0  \}\cr
 &=\{x \in \R^d: 
 x_d>  \max_{i:i\neq j} \frac{ \langle x', B'_i -B'_j \rangle }{B_{j,d}-B_{i,d}} \}
 \neq \emptyset\cr
\end{split} 
\end{equation}
$\Cox$

Before we state our theorem let us introduce the kernels 
\begin{equation}
\begin{split}\label{colonel}\g[b](a| \nu)=\frac{e^{-d F_{\nu}(a)}\a[b](a)}{
\sum_{\bar a\in E}   e^{-d F_{\nu}(\bar a)}\a[b](\bar a)}
\end{split} 
\end{equation}
with $\nu \in \PP(E)$. These are the limiting local distributions 
of a spin at a site with a disorder variable in the state $b$ if the empirical spin-average of the rest 
of the system is given by the measure $\nu$. The products over all sites 
of these quantities, for $\nu=\pi \hat \nu_j$, 
will play the role of pure measures.

We are now in the position to give our main result. 

\begin{thm} \label{thm1}Assume that the model satisfies the {\em non-degeneracy assumptions 1)} and {\em 2)}. 
Define the weights 
\begin{equation}
\begin{split}\label{nondeg4}
&w_j:=\P_{\pi}(G\in R_j) 
\end{split} 
\end{equation}
where $G\in  T\PP(E')$ is a centered Gaussian variable with the same covariance as\\$\sqrt{n}(\hat\pi_n-\pi)$ which is given 
by the expression $ C_{\pi}(b,b')=\pi(b)1_{b=b'}-\pi(b)\pi(b')$. 

Then $\sum_{j=1}^k w_j=1$ and  the metastate on the level of the states equals 
\begin{equation}
\begin{split}\label{AW-metastate}
&\k[\eta](d\mu)=\sum_{j=1}^k w_j \d_{\mu_j[\eta]}(d\mu)
\end{split} 
\end{equation}
where 
$\mu_j[\eta]:= \prod_{i=1}^\infty \g[\eta(i)](\,\cdot\,| \pi \hat\nu_j)$. 
 
\end{thm}

{\bf Comment. } We like to reformulate our result on the visibility or 
invisibility of the phases in the following way. 
Let us denote by  $M^{**}=\{\hat\nu \in M^* : w_{\hat \nu}>0\}$ 
the subset of {\em visible pure phases} in the pure phases $M^*$. Let us use 
the symbol $B_{\cdot}$ for the bijection (under our hypothesis)  
\begin{equation*}\begin{split}
B_{\cdot}:M^*  &\rightarrow T \PP(E') \cr
\hat \nu&\mapsto  B_{\hat \nu}
\end{split}
\end{equation*} 
Then we can write in short 
\begin{equation*}\begin{split}
M^{**}=(B_{\cdot})^{-1}\Bigl(\text{ex}( \HH_{\text{conv}}(B_{\cdot}(M^*) )    \Bigr)
\end{split}
\end{equation*}

Let us derive the following immediate consequence 
which provides a a
symmetry, due to the randomness (the symmetry of the Gaussian, obtained
via the CLT).  
 
\begin{cor} Suppose that the system admits precisely two pure phases, i.e. $|M^*|=2$. 
 Then the metastate is the symmetric mixture between the two, i.e. 
 \begin{equation}
\begin{split}\label{AW-metastate}
&\k[\eta](d\mu)=\frac{1}{2} \d_{\mu_1[\eta]}(d\mu)+\frac{1}{2} \d_{\mu_1[\eta]}(d\mu)
\end{split} 
\end{equation}
\end{cor}
The corollary is clear from the theorem since in that case $R_1=-R_2 $ 
and this implies by the {\em non-degeneracy assumption 2)} that $w_1=w_2$.

\begin{cor} Suppose that the random-field is two-valued, i.e. $|E'|=2$, and the number 
of pure phases $|M^*|\geq 2$ arbitrary.   Then the set of visible states has 
two elements  and $w(\hat \nu)=\frac{1}{2}$ for both elements $\hat\nu\in M^{**}$. \end{cor}
The corollary 
is clear from the theorem since any convex polyhedron in one dimension has only 
two extremal points. 

For illustrational purposes recall the situation in  the mean-field random field Ising 
model with two-valued symmetrically distributed random field with coupling strength $\e$ 
and temperature $\b^{-1}$. In this model the $\b^{-1},\e$-plane contains a bounded open region 
for which $|M^{*}(\b^{-1},\e)|=2$. The boundary of this region is a curve which 
splits into a part for which $|M^{*}(\b^{-1},\e)|=3$ and a part for which 
$|M^{*}(\b^{-1},\e)|=1$. In the complement of the union of those previous regions 
we have $|M^{*}(\b^{-1},\e)|=1$. 
The situation of the second corollary is met on the curve where $|M^{*}(\b^{-1},\e)|=3$.

\subsubsection{Exploiting the mean-field equation}

Using variational calculus and assuming 
differentiability of $F$ one sees that the minimizers 
of the variational problem above must satisfy the consistency (mean-field) equations 
\begin{equation}
\begin{split}\label{1.3.2}
&\hat \nu[b](a)=\g[b](a| \pi \cdot \hat \nu)
\end{split} 
\end{equation}
which are coupled over $b\in E'$. Summing over these indices 
one gets the mean-field equation for the total empirical mean $\nu= \pi \cdot \hat \nu$ 
of the form  
\begin{equation}
\begin{split}\label{1.3.2}
&\nu(a)=\sum_{b\in E'}\pi(b)\g[b](a|\nu)
\end{split} 
\end{equation}
We note the following Lemma.  

\begin{lem} Define the function $\hat\G: \PP(E)\rightarrow \PP(E)^{E'}$ by the r.h.s. of the mean field equation, namely
\begin{equation}
\begin{split}\label{1.3.3}
\hat \G(\nu)=\Bigl(\g[b](\cdot | \nu)\Bigr)_{b\in E'}
\end{split} 
\end{equation}
Define the function $\hat B: \PP(E)\rightarrow T\PP(E')$ by 
\begin{equation}
\begin{split}\label{1.3.3-0}
\hat B_{\nu}[b]&= \log \sum_{a\in E}e^{-dF_{\nu}(a)}\a[b](a)- C \cr 
C&= \frac{1}{|E'|}\sum_{b\in E'}\log \sum_{a\in E}e^{-dF_{\nu}(a)}\a[b](a)
\end{split} 
\end{equation}
Then, for all $\hat \nu \in M^*$ we have that 
\begin{equation}
\begin{split}\label{1.3.3+0.1}
\hat\nu&=\hat \G(\pi \hat \nu) \cr
B_{\hat\nu}&=\hat B_{\pi \hat \nu}\cr
\end{split} 
\end{equation}
For all $\nu \in \pi M^*$ we have that the free energy can be written as 
\begin{equation}
\begin{split}\label{1.3.3+0.2}
\Phi[\pi](\hat \G(\nu))&=F(\nu)-\langle d F_{\nu},\nu\rangle-\langle \hat B_{\nu},\pi\rangle + C
\end{split} 
\end{equation}
\end{lem}
The first statement is just a rephrasing of the mean-field equation. It serves us 
to see that there is a bijection between $\pi M^*=\{\pi \hat \nu| \hat \nu \in M^*\}  \subset  \PP(E)$ (a subset in a space of measures 
with dimension $|E|-1$) and $M^*$ (a subset in a space of measures 
with dimension $(|E|-1)^{|E'|}$) .  
The second part means that the logarithm of the normalization factor ("little partition function") 
of the mean-field kernels in the total empirical distribution $\nu$ 
of type $b$ produces the $b$'th component of the stability vector 
corresponding to the minimizer with total empirical mean $\nu$.  

The interesting feature is that the form of $\pi$ does not enter 
at all into this formula (it enters however through the question 
which minimizer $\hat\nu$ and hence also $\nu$ appears.) 

{\bf Proof.} The first part is obvious. To prove the second part, 
for $\hat \nu \in M^*$ we write, with a constant $C'$ to be determined 
\begin{equation}
\begin{split}\label{1.3.3+0}
-B_{\hat\nu}[b]&=\sum_{a\in E}d F_{\pi\cdot \hat \nu}(a)\hat\nu[b](a)+S(\hat \nu[b]|\a[b])-C'\cr 
&\equiv \hat\nu[b]( d F_{\pi\cdot \hat \nu}(\cdot) )+S(\hat \nu[b]|\a[b])-C'\cr 
&=-\log \sum_{a\in E}e^{-dF_{\pi\hat \nu}(a)}\a[b](a)- C' \cr
\end{split} 
\end{equation}
where the last equality follows from the mean-field equation. 
This proves the second claim. The last claim follows from the first equality of the last display multiplying with 
$\pi(b)$ and summing 
over $b\in E'$.  
$\Cox$


\subsection{Ising random-field examples} 

Let us take the Ising model with $F(\nu)=- \b (\nu(+)^2+ \nu(-)^2) $. 

Any possible local single-site measure $\a$ can be described as an $\a[h](\s_i)= \frac{e^{h \s_i}}{2 \cosh h}$. 
Any $\nu=\nu_m$ can be described in terms of its mean value $\nu_m(+)-\nu_m(-)=m$. 

So we can write 
\begin{equation}
\begin{split}\label{1.3.3}
\hat B_{\nu_m}[h]\equiv  \hat B_{\nu_m}[\a[h]]
&= \log \frac{e^{\b 2\frac{1+m}{2}+ h}+ e^{\b 2 \frac{1-m}{2}- h}}{2 \cosh h}-C\cr
&= \b +\log\frac{\cosh (\b m  +h)}{ \cosh h} - C \cr 
\end{split} 
\end{equation}

Let us now fix $E'=\text{supp}(\pi)=\{ \a_h: h\in \{h_1, h_2, \dots, h_L \}   \} $ as 
the set of allowed local measures.  This gives us the normalized vector 
in the tangent space $T \PP(E')$ with entries 
\begin{equation}
\begin{split}\label{1.3.3}
\hat B_{\nu_m}[h_i]:= \log\frac{\cosh (\b m  +h_i)}{ \cosh h_i} - \frac{1}{L}\sum_{j=1}^L \log\frac{\cosh (\b m  +h_j)}{ \cosh h_j} \cr
\end{split} 
\end{equation}
Writing a vector with $L=|E'|$ components we have 
\begin{equation*}\begin{split}
\hat B_{\nu_m}=\begin{pmatrix}
  \log\frac{\cosh (\b m  +h_1)}{ \cosh h_1} \\
  \dots \\
  \log\frac{\cosh (\b m  +h_L)}{ \cosh h_L} 
\end{pmatrix}- \frac{1}{L}\sum_{j=1}^L\log\frac{\cosh (\b m  +h_j)}{ \cosh h_j}
\begin{pmatrix}
 1\\
  \dots \\
 1
 \end{pmatrix}
\end{split}
\end{equation*}

\begin{lem} Let $E'\sb \R$, $2\leq |E'| <\infty$. 
Then the map $m \mapsto \hat B_{\nu_m}$ is injective. 
\end{lem}

{\bf Proof. } We have at least two elements,  $h_1<h_2$ (after possible change of indices) in $E'$. 
Let $\nu_m,\nu_{\tilde m}$ be given with $\hat B_{\nu_m}=\hat B_{\nu_{\tilde m}}$. 
By easy manipulations looking at the first two components 
of $B$ the latter implies that 
\begin{equation}\begin{split}\label{2323}
\frac{\cosh (\b m  +h_1)}{  \cosh (\b m  +h_2)}=
\frac{  \cosh (\b \tilde m  +h_1)}{  \cosh (\b \tilde m  +h_2)}
\end{split}
\end{equation}
From this follows $m=\tilde m$ by injectivity of the function $x\mapsto \frac{\cosh x}{\cosh (x+1)}$. \\
$\Cox$




Let us extend the random-field Ising model to a non-quadratic Hamiltonian $F(\nu)=G(\nu(+)-\nu(-))$ and 
general local measures $\a=(\a[h])_{h\in E'}$ with a finite set $E'$ just as above in the quadratic case. 

Then the mean field equation becomes 
\begin{equation}
\begin{split}\label{1.3.2}
&m=\sum_{i=1}^L \pi(h_i)\tanh(-G'(m)+h_i)\cr
\end{split} 
\end{equation}
The stability vector becomes 
\begin{equation}
\begin{split}\label{1.3.3}
\hat B_{\nu_m}[h_i]:= \log\frac{\cosh (-G'(m)  +h_i)}{ \cosh h_i} - \frac{1}{L}\sum_{j=1}^L \log\frac{\cosh (-G'( m)
  +h_j)}{ \cosh h_j} \cr
\end{split} 
\end{equation}
Then the injectivity of the map $m\mapsto \hat B_{\nu_m}$ holds under the assumption 
that $m\mapsto G'(m)$ is injective, by the same proof, replacing 
$m$ by $-G'(m)$ in  \eqref{2323}. 

We have thus proved the following statement. 

\begin{prop} For a random-field Ising model with Hamiltonian $F(\nu)=G(\nu(+)-\nu(-))$ and $G'$ injective 
the second non-degeneracy assumption is automatically satisfied, for any distribution of random 
fields with finite support.  
\end{prop}

It is easy to create a two-minima situation where there is no symmetry, by looking 
at the equal-depth condition for the free energy 
\begin{equation*}
\begin{split}\label{1.3.2}
&\Phi[\pi](\hat \G(\nu_m))=F(\nu_m)-\sum_{a\in E}dF_{\nu_m}(a)\nu_m(a)- \sum_{b\in E'}\pi(b)\log \sum_{a\in E}e^{-dF_{\nu_m}(a)}\a[b](a)\cr
&=G(m)-mG'(m)-\sum_{i=1}^L \pi(h_i)
\log\frac{\cosh (-G'(m)  +h_i)}{ \cosh h_i}
\end{split} 
\end{equation*}
where both minima would get the same weight in the metastate necessarily. 

In fact, a situation with precisely two minimizers not related by symmetry 
was proved to occur (even) for the (symmetric) model $G(m)=-\frac{\b m^2}{2}$, $E=E'=\{1,-1\}$, 
$\pi(1)=\frac{1+\a}{2}=1- \pi(-1)$, $\a[b](a)=\frac{e^{\b \e a b}}{\cosh \b \e}$, for the region $R_{\text{34}}$ in 
the $(\b^{-1},\e)$-plane characterized in \cite{KuLN07} and depicted below,  
for a suitable choice of $\a=\a(\b,\e)>0$.  

\begin{center}
\includegraphics[height=8cm]{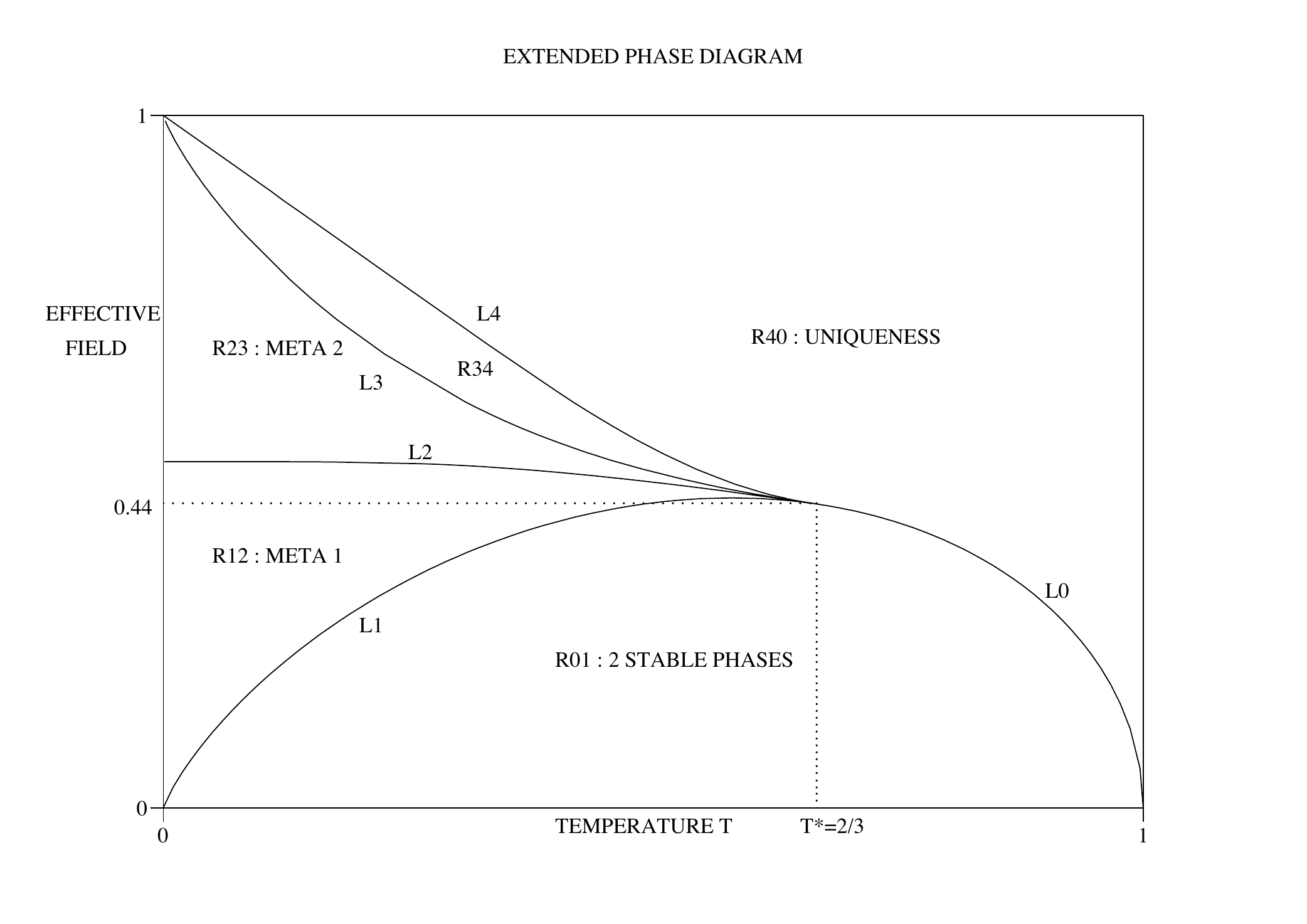}
\end{center}

\subsection{Potts random field examples}

Let us take the Potts model with  quadratic interaction 
$$F(\nu)=- \frac{\b}{2} (\nu(1)^2+\dots+ \nu(q)^2)$$  
in the presence of 
the local single-site measures $\a[b](\s_i)$ (specified below) where we write  
$$E'=\text{supp}(\pi)=\{\a[b]: b\in \{b_1, b_2, \dots, b_L \}   \} $$ 
Then we have for the stability vector 
\begin{equation*}\begin{split}
\hat B_{\nu}=\begin{pmatrix}
  \log \sum_{a=1}^q e^{\b \nu(a)}\a[b_1](a) 
    \\
  \dots \\
  \log \sum_{a=1}^q e^{\b \nu(a)}\a[b_L](a) 
\end{pmatrix}- \frac{1}{L}\sum_{j=1}^L\log \sum_{a=1}^q e^{\b \nu(a)}\a[b_j](a)
\begin{pmatrix}
 1\\
  \dots \\
 1
 \end{pmatrix}
\end{split}
\end{equation*}
{\bf Remark. } The map $\hat B_{\cdot }: \PP(E)\rightarrow T \PP(E')$ is a map 
between spaces of dimension $|E|-1$ and $|E'|-1$. It has a chance to 
be injective as such (on the whole space $\PP(E)$) only when $|E'|\geq |E|$. 

Let us take $E\equiv E'$ and $\pi$ to be the equidistribution and 
switch to the specific case $\a[b](a)=\frac{e^{B 1_{b=a}}}{e^B + q-1}$
(random field with homogenous intensity). 
The kernels become 
$$
\g[b](a| \nu)=\frac{e^{\b \nu(a) + B 1_{a=b}}}{
\sum_{\bar a\in E}  e^{\b \nu(\bar a) + B 1_{\bar a=b}}}
$$
We will be looking at measures in $\nu_{j,u}\in \PP(E)$ of the form 
$\nu_{j,u}(j)=\frac{1+ u (q-1)}{q}$,  $\nu_{j,u}(i)=\frac{1- u}{q}$ 
for $i\neq j$. 
The stability vector for $\nu_{1,u}$ is given by
\begin{equation*}\begin{split}
\hat B_{\nu_{1,u}}
= \begin{pmatrix}
  \frac{q-1}{q}\log\frac{e^{\b u + B} + q-1}{e^{\b u}  + e^{B} + q-2}\\
   -\frac{1}{q}\log\frac{e^{\b u + B} + q-1}{e^{\b u}  + e^{B} + q-2}\\ 
  \dots \\ 
  -\frac{1}{q}\log\frac{e^{\b u + B} + q-1}{e^{\b u}  + e^{B} + q-2}
\end{pmatrix}\cr
 \end{split}
\end{equation*}
the other ones are related by symmetry. 
We note that the first entry is strictly positive while the other entries are negative 
(for $B>0$ and $u>0$). 

We have the mean-field equation for $u$ of the form 
%
\begin{equation}
\begin{split}\label{rr}
 u &=\frac{e^{\b u}}{ 
e^{\b u} + e^B + (q-2)}
- \frac{1}{
e^{\b u + B} +  (q-1)}
\end{split} 
\end{equation}\\
We notice that $u=0$ is always a solution, and for $B=0$ we obtain exactly the known 
mean-field equation for Potts without disorder. The latter model shows 
a first-order transition as a function of temperature 
at critical temperature $\b_c=\frac{2(q-1)}{q-2}\log (q-1)$ 
\cite{ElWa90}. \\
The r.h.s. of \eqref{rr} is always positive, as a computation shows.  This gives rise to 
a non-trivial solution $u$, in a certain range of parameters. Note that this non-trivial 
solution is not always the one to be chosen. It is to be chosen iff 
$\Phi[\pi](\hat \G(\nu_{j,u}))<
\Phi[\pi](\hat \G(\nu_{j,u=0}))
$. So, the first order transition point is given by equality 
of the last equation. 
%
Forgetting a $u$-independent term we have, independently of the direction $j$,  
\begin{equation}
\begin{split}\label{1.3.2}
&\Phi[\pi](\hat \G(\nu_{j,u}))= 
\log\frac{e^B+q-1}{e^{\b u}+e^B+q-2}+\frac{\b(q-1)}{2 q}u^2+\frac{\b}{q}u -\frac{1}{q}\log\frac{e^{\b u+B}+q-1}{e^{\b u}+e^B+q-2}
\end{split} 
\end{equation}
with the property that $\Phi[\pi](\hat \G(\nu_{j,u=0})=0$. 
For illustrational purposes 
let us focus on the case $q=3$. We don't provide 
a complete bifurcation analysis here, but just outline the picture. 
The case $B=0$ is perfectly 
understood and we know that there is a first order transition at the critical inverse temperature
$\b= 4 \log 2$. The nature of the transition stays the same when $B$ takes small enough positive values 
and there is a line in the space of temperature and coupling strength $B$ 
of an equal-depth minimum at $u=0$ and a positive value of $u=u^*(\b,q)$. (See Fig. 1 for a numerical example.)  
Along this line the set of Gibbs measures is strictly bigger then the set of states which 
are seen under the metastate. 

The Plot shows the graph of $u\mapsto\Phi[\pi](\hat \G(\nu_{j,u}))$
for $B=0.3, q= 3, \b= 4 \log 2  + 0.03203$ at which there is the first order transition. 

\begin{center}
\includegraphics[height=6.0cm]{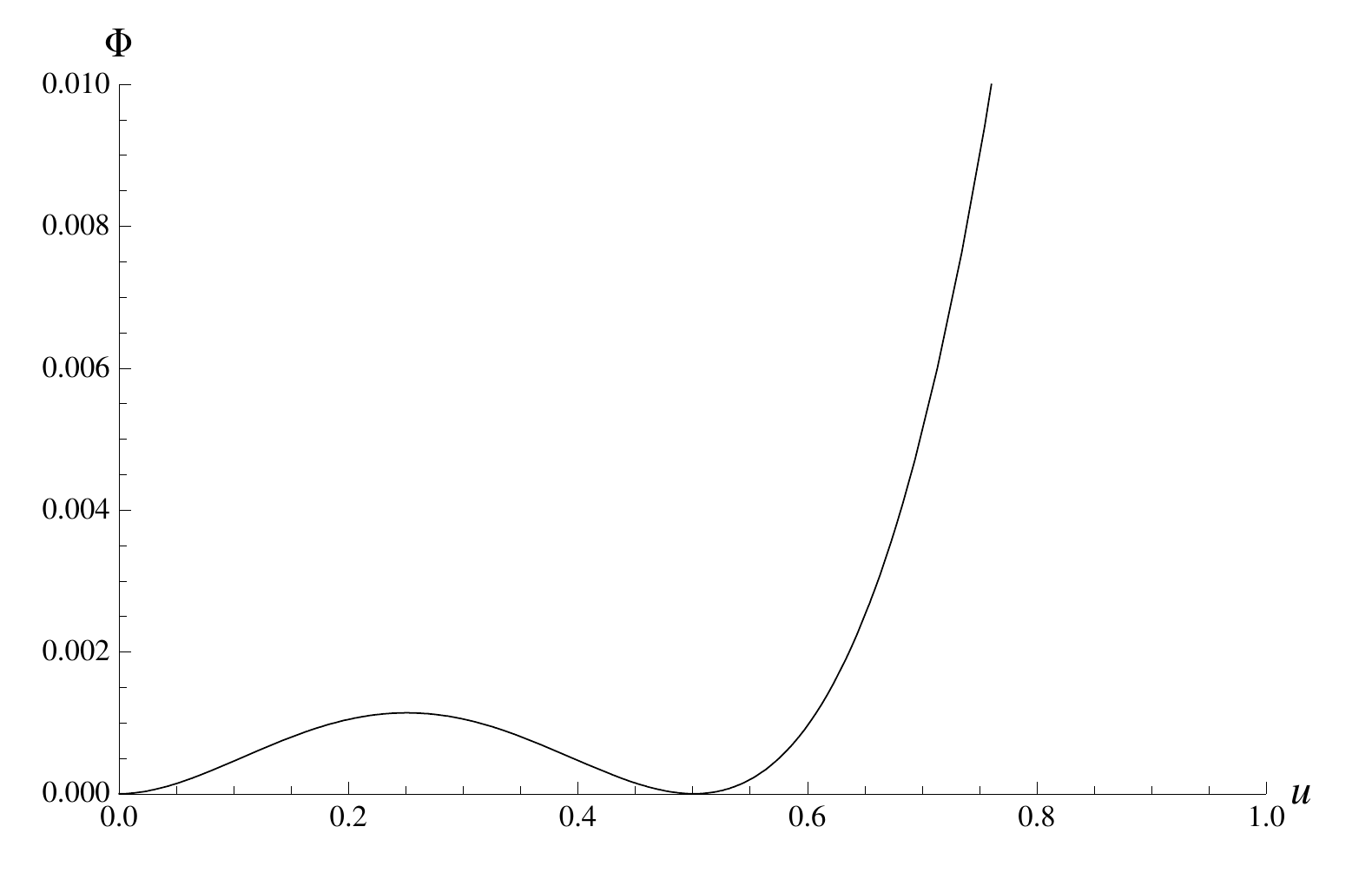}
\end{center}

The metastate becomes 
$\k[\eta](d\mu)=\frac{1}{3}\sum_{j=1}^3 \d_{\mu_j[\eta]}$ 
with  $\mu_j[\eta]= \prod_{i=1}^\infty \g[\eta(i)](\,\cdot\,| \nu_{j,u=u^*(\b,q)})$. 
This follows from the form of the stability vector using that $\hat B_{\nu_{1,u=0}}=0$ 
and hence lies in the convex hull of the three others. 
\bigskip
\bigskip

\subsection{Strategy of proof, non-degeneracy assumption and concentration}

The outline of the remainder of the paper is as follows. 

We begin in Section 2 with a discussion of another related result, namely 
the metastate on the level of the empirical spin-distributions. 
The theorem is 
quite analogous, the same weights $w_j$ appear, and the proof is slightly 
easier than that of the full theorem. To arrive at the proof of this theorem we will discuss the concentration property 
of the vector of the empirical 
distributions for good realizations of the disorder which 
will force the system to be in one definite state. 
In particular it will show how the non-degeneracy 
assumptions 1) and 2) are naturally used in that argument and this will explain how the CLT for 
empirical distributions of disorder variables translates into the form of the weights $w_j$.  

Then we will turn in section 3 to the proof of the metastate theorem on the level of states 
and conclude. 
\bigskip
\bigskip 

\section{The Metastate on the level of the \\ empirical spin-distribution}

Two ways of looking at the spin-distributions of disordered mean-field systems 
are natural. In the first one, described in the introduction, we focus on measures of the spins 
themselves, and evaluate them on local observables. In the second one, we focus on aggregate properties 
of the system, and look at functions of the empirical spin-distribution of the whole system. 
From the second point of view it is natural to make the following definition of a metastate on the level 
of the empirical spin-distribution.

Denote by $\r[\eta](n):=\mu_{F,n}[\eta](L_n)$ the image of the finite-volume Gibbs-measure under 
the empirical distribution. This defines a disorder-dependent 
element in $\PP(\PP(E))$. Under our assumptions these measures will concentrate 
on the finite set $\pi M^*=\{\pi \hat \nu_j, j=1,\dots,k\}$. It is useful to introduce a metastate 
which tells us more precisely how this concentration takes place. This is the reason 
for the following definition.

\begin{defn} Assume that, for every bounded continuous $\Psi:\PP(\PP(E))\times (E')^\infty$  
the limit 
\begin{equation}
\begin{split}\label{nondeg3}
&\lim_{n\uparrow \infty}\int\P(d\eta)\Psi(\r[\eta](n),\eta)=\int K(d\r,d\eta)\Psi(\r,\eta)\cr
\end{split} 
\end{equation}
exists. Then the conditional distribution $\bar \k[\eta](d\r):=  K(d\r|\eta)$ 
is called the {\em metastate on the level of the empirical spin-distribution}. 
\end{defn}

Believing in the first theorem it is not surprising that this metastate takes the following form. 

\begin{thm} \label{thm2}Under the non-degeneracy assumptions 1) and 2), we have 
\begin{equation}
\begin{split}\label{AW-metastate}
&\bar\k[\eta](d\r)=\sum_{j=1}^k w_j \d_{\d_{\pi\hat\nu_j}}(d\r)
\end{split} 
\end{equation}
for $\P_{\pi}$-a.e. $\eta$. 
\end{thm}

As a difference with respect to the first theorem let us point out that in this case the dependence 
on the disorder has vanished on the r.h.s. 
\bigskip
\bigskip

{\bf Proof of Theorem \ref{thm2}.}
For $n_1<n_2$ integers, let's define
\begin{equation}\label{X}
 X_{[n_1,n_2]}[\eta]=\frac{1}{\sqrt{n_2-n_1+1}}\sum_{i=n_1}^{n_2}\d_{\eta_i} - \sqrt{n_2-n_1+1} \; \pi
\end{equation}

Define $n,l$-dependent {\em good-sets} $\HH_{n,l}^{\t}$ of the realization of the randomness as follows
\begin{equation}
\begin{split}\label{nondeg3}
&\HH_{i,n,l}^{\t}:=\left\lbrace \eta \in (E')^{n-l}: X_{[l+1,n]}[\eta]\in R_{i,n}^{\t}\right\rbrace  \cr 
&\HH_{n,l}^{\t}:=\bigcup_{i=1}^k \HH_{i,n,l}^{\t}
\end{split} 
\end{equation} 
where $R_{i,n}^{\t}:=\{x\in T\PP(E'): \langle x,  B_i \rangle-\max_{k\neq i}    \langle x, B_k \rangle> n^{-\frac{1}{2}+\t}, \Vert x\Vert \leq n^{\frac{\t}{4}} \}$, where $0<\t<\frac{1}{2}$. For the sake of clarity set $\d_n= n^{-\frac{1}{2}+\t}$. The chosen range of $\t$ ensures that 
$\d_n\downarrow 0$, but not too fast, namely in  such a way 
that $\sqrt{n}\;\d_n\uparrow\infty$.\\
$\HH_{i,n,l}^{\t}$ is a region of the disorder random variables 
which allows us to deduce that the measure on the empirical distribution will 
be with large probability inside a ball around $\pi \hat\nu^*_i$.
\vskip .1in
\noindent
{\bf Remark:} We need $\d_n\downarrow 0$ because we want to cover all of the corresponding {\em stability-region $R_i$} \eqref{RJ} in the large-$n$ limit.
The condition regarding the velocity with which $\d_n$ is going to $0$ ensures the concentration of the measure around a particular minimizer, in other words it will enable us to see the breaking of the degeneracy of the minimizers caused by the fluctuations of $\hat\pi_n$. The relevance of the cutoff $\Vert x\Vert \leq n^{\frac{\t}{4}}$ will be seen later. 
\bigskip 

\noindent
\vskip .1in
\noindent


\begin{lem}\label{concentrationH}
Let us assume that $\eta\in \HH_{i,n,0}^{\t}$. Then 
\begin{equation}
\begin{split}\label{lemmaball}
&\mu_{F,n}[\eta(1),\dots,\eta(n)](d( L_n,\pi \hat\nu^*_i) \leq \e)\geq 1 - \bar r(\e,n)\cr
\end{split} 
\end{equation}
where $\lim_{n\uparrow\infty}\bar r(\e,n)=0$ for all $\e>0$.

\end{lem}

{\bf Proof: } Call $M_n :=\{\nu \in \PP(E): \exists \o\in E^n \text{ such that } L_n^\o=\nu\}$. To every element $\nu \in M_n$ correspond several possible values of the empirical distribution vectors $\hat L_n\in \PP(E)^{E'}$, given $\hat\pi_n$. We call this set
$\hat M_n:=\{\hat\nu \in \PP(E)^{E'}:\exists \o\in E^n \text{ such that }\hat\nu=\hat L_n \}$.
Let's define $\r^{\e}[\eta](n) \in \PP (\pi M^* )$ assigning probability weights to the $\e$-balls by  
\begin{equation}
\begin{split}
&\r^{\e}[\eta](n)(\pi\hat\nu_i^*):=\frac{\mu_{F,n}[\eta(1),\dots,\eta(n)]( L_n \in B(\e,\pi\hat \nu^*_i))}{\displaystyle\sum_{j=1}^k \mu_{F,n}[\eta(1),\dots,\eta(n)]( L_n \in B(\e,\pi\hat \nu^*_j))}\cr
\end{split}
\end{equation}
At this stage the measures appearing in the former definition involve a sum over $\nu \in M_n\bigcap B(\e,\pi\hat\nu_i^*)$ and for the correspondence formerly mentioned we can write
\begin{equation}
\begin{split}
 \r^{\e}[\eta](n)(\pi\hat\nu_i^*)=&\frac{\displaystyle\sum_{\hat \nu:\hat\pi_n \hat \nu \in B(\e,\pi\hat\nu^*_i)}\mu_{F,n}[\eta(1),\dots,\eta(n)]( \hat L_n = \hat \nu)}
{\sum_{j=1}^k\displaystyle\sum_{\hat \nu:\hat\pi_n \hat \nu \in B(\e,\pi\hat\nu^*_j)}\mu_{F,n}[\eta(1),\dots,\eta(n)](\hat L_n = \hat \nu)}\cr
=&\frac{\displaystyle\sum_{\hat \nu:\hat\pi_n \hat \nu \in B(\e,\pi\hat\nu^*_i)}\frac{\displaystyle\sum_{\sigma \in E^n : \hat L_n^\sigma=\hat\nu}e^{-nF(\hat\pi_n\hat\nu)}\prod_{i=1}^n\alpha[\eta_i](\sigma_i)}
{
\displaystyle\sum_{\bar\nu\in \hat M_n} \sum_{\sigma \in E^n : \hat L_n^\sigma=\bar \nu}e^{-nF(\hat\pi_n \bar\nu)}\prod_{i=1}^n\alpha[\eta_i](\bar{\sigma_i})}}
{\displaystyle\sum_{j=1}^k \sum_{{\hat \nu:}\atop{\hat\pi_n \hat \nu \in B(\e,\pi\hat\nu^*_j)}}\frac{\displaystyle\sum_{\sigma \in E^n : \hat L_n^\sigma=\hat\nu}e^{-nF(\hat\pi_n\hat\nu)}\prod_{i=1}^n\alpha[\eta_i](\sigma_i)}
{
\displaystyle\sum_{\bar\nu\in \hat M_n} \sum_{\sigma \in E^n : \hat L_n^\sigma=\bar\nu}e^{-nF(\hat\pi_n \bar\nu)}\prod_{i=1}^n\alpha[\eta_i](\bar{\sigma_i})}}\cr
\end{split}
\end{equation}

Decomposing the spin-sums into 
sums over possible values of the vector of empirical distributions on the $b$-like sites we can rewrite 
the last expression  as
%
\begin{equation}
\begin{split}\label{lemmRho}
&\frac{\displaystyle\sum_{\hat \nu:\hat\pi_n \hat \nu \in B(\e,\pi\hat\nu^*_i)}e^{-nF(\hat\pi_n\hat\nu)}\prod_{b=1}^{\mid E'\mid}\alpha[b]^{\mid \Lambda_n(b) \mid}(\Omega_{\mid \Lambda_n(b)\mid}(\hat\nu(b)))}
{\displaystyle\sum_{j=1}^k \sum_{\hat \nu:\hat\pi_n \hat \nu \in B(\e,\pi\hat\nu^*_j)}e^{-nF(\hat\pi_n\hat\nu)}\prod_{b=1}^{\mid E'\mid}\alpha[b]^{\mid \Lambda_n(b) \mid}(\Omega_{\mid \Lambda_n(b)\mid}(\hat\nu(b)))}\cr
\end{split}
\end{equation}
where $  \Omega_{\mid \Lambda_n(b)\mid}(\hat\nu(b)) = \{ \sigma \in E^{\mid \Lambda_n(b)\mid}: \hat L_n^{\sigma}(b)= \hat \nu(b) \}  $ , and $\alpha[b]^{\mid \Lambda_n(b) \mid}(\cdot)$ is the product measure on the $b-$like sites. For sake of clarity let us recall the finite
volume finite alphabet version of Sanov's theorem which is stated as Lemma 2.1.8 in \cite{DeZe10}  
which we will make use of in the next step.
\begin{lem} 
Let $\nu$ be a probability measure on a finite state space $E$. 
For fixed $n$ define the set of {\em microstates} compatible with $\nu$ by  
\begin{equation}
 \O(\nu):=\{\o \in E^n |L_n^\o=\nu \}
\end{equation}
Then, if $n\nu(x)$ is integer-valued for all $x \in E$ we have the upper and lower large deviation bounds
\begin{equation}\label{DZ}
 (n+1)^{-|E|}e^{-nS(\nu|\mu)}\leq \mu(\O(\nu))=\mu(\{\o:L_n^\o=\nu \})\leq e^{-nS(\nu|\mu)}
\end{equation}
 
\end{lem}
Using \eqref{DZ} we get the lower bound to \eqref{lemmRho} of the form 
\begin{equation}
\begin{split}\label{rho2}
&\frac{1}{1+\displaystyle\sum_{j\neq i}^{k}\frac{\displaystyle\sum_{\hat \nu:\hat\pi_n \hat \nu \in B(\e,\pi\hat\nu^*_j)}e^{-n\Phi[\hat\pi_n](\hat\nu)}}{\displaystyle\sum_{\hat \nu:\hat\pi_n \hat \nu \in B(\e,\pi\hat\nu^*_i)}e^{-n\Phi[\hat\pi_n](\hat\nu)}\prod_{b\in E'}(\mid \Lambda_n(b)\mid +1)^{-\mid E'\mid}}}\cr
\end{split}
\end{equation}
Let us notice that, in the last expression, the free energy \eqref{fi} has appeared. However it does not involve yet the minimizer $\hat\nu_i^*$ in an explicit way.
What we would like to do next, is to understand the $\hat \pi_n$-dependence of the minima in the different balls. Differences 
in the depths of the minima would not be present for $\hat\pi_n=\pi$ but will be created  by the fluctuations of $\hat\pi_n$.  


In order to achieve this we first need to compare the values that the $\hat\pi_n-$de\-pen\-dent free-energy takes on the ball with the one corresponding to the center. 
 As we will see in Proposition \ref{pp2.5} this can be done uniformly with respect to the centers ($\pi-$minimizers). Secondly we will compare, for any fixed minimizer, the difference between the $\hat\pi_n-$dependent free-energy and the $\pi-$dependent one; this will be done using the linearization procedure \eqref{li-fi}. Let us emphasize the fact that the definition of the {\em good-sets} $\HH_{i,n,l}^{\t}$ has been chosen {\em ad hoc} to guarantee, in the limit $n \uparrow\infty$, that the $i$-th stability vector will "dominate" the others, and thus the concentration around $\hat\nu_i^*$ will take place. We also need an upper bound on $\Vert\hat\pi_n-\pi\Vert$ for that procedure to work which is the reason 
 for the cutoff in the definition of the {\em good-sets}. 
 
 The next proposition formalizes the first step.
\begin{prop}\label{pp2.5}
 Under the non-degeneracy assumption 1) there exists an $\e_0>0$ and 
 a positive constant $K$ such that for all $\e\leq \e_0$ and
 for $n$ sufficiently large
\begin{equation}\label{inf-bound}
 \frac{-K\Vert\hat\pi_n-\pi\Vert^2}{2}\leq \inf_{\hat\nu \in \hat M_n \bigcap B(\e,\pi\hat\nu_j^*)}(\Phi[\hat\pi_n](\hat\nu)-\Phi[\hat\pi_n](\hat\nu_j^*))
\end{equation}
for all minimizers $\hat\nu_j^*$.
\end{prop}

{\bf Proof: } We will show that there exists a positive $K$ such that
\begin{equation}
 \frac{-K||\hat\pi_n-\pi||^2}{2}\leq \inf_{\hat\nu \in B(\e,\pi\hat\nu_j^*)}(\Phi[\hat\pi_n](\hat\nu)-\Phi[\hat\pi_n](\hat\nu_j^*))
\end{equation}
holds, hence the proposition will follow using the inequality $\displaystyle \inf_{\hat\nu \in \hat M_n \bigcap B(\e,\pi\hat\nu_j^*)}\leq \inf_{\hat\nu \in \hat B(\e,\pi\hat\nu_j^*)}$.\\
Let us take a Taylor expansion of $\Phi[\hat\pi_n](\hat\nu)$ around $\hat\nu^*_j$, namely
\begin{equation}
\begin{split}
  &\Phi[\hat\pi_n](\hat\nu)=\cr 
&\Phi[\hat\pi_n](\hat\nu^*_j)+\langle \nabla\Phi[\hat\pi_n](\hat\nu^*_j),\hat\nu-\hat\nu^*_j \rangle+\frac{1}{2}\langle\hat\nu-\hat\nu^*_j,H\Phi[\hat\pi_n](\hat\nu^*_j)(\hat\nu-\hat\nu^*_j)\rangle+||\hat\nu-\hat\nu^*_j ||^2 R(\hat\nu,\hat\nu^*_j)\cr
\end{split}
\end{equation}
where $R(\hat\nu,\hat\nu^*_j)$ is a continuous function at $\hat\nu=\hat\nu^*_j$ with $R(\hat\nu^*_j,\hat\nu^*_j)=0$, and $H$ is the Hessian.

So we have
\begin{equation}
 \begin{split}
&\Phi[\hat\pi_n](\hat\nu)-\Phi[\hat\pi_n](\hat\nu^*_j)\geq \langle \nabla\Phi[\hat\pi_n](\hat\nu^*_j),\hat\nu-\hat\nu^*_j \rangle+\frac{1}{2}\langle\hat\nu-\hat\nu^*_j,(H\Phi[\hat\pi_n](\hat\nu^*_j)-C_1)(\hat\nu-\hat\nu^*_j)\rangle\cr
 \end{split}
\end{equation}
where $C_1$ is a non negative constant which can be chosen arbitrarily close to zero when we restrict to balls with sufficiently small radii $\e$.

The $\inf$ for the previous r.h.s. is obtained at the point
$\hat\nu^*=-(H\Phi[\hat\pi_n](\hat\nu^*_j)-C_1)^{-1}\nabla\Phi[\hat\pi_n](\hat\nu^*_j)+\hat\nu^*_j$.\\
So we have
\begin{equation}
\begin{split}
&\displaystyle\inf_{\hat\nu\in B(\e,\pi\hat\nu_j^*)}(\Phi[\hat\pi_n](\hat\nu)-\Phi[\hat\pi_n](\hat\nu_j^*))\geq  -\frac{1}{2}\langle\nabla\Phi[\hat\pi_n](\hat\nu^*_j),(H\Phi[\hat\pi_n](\hat\nu^*_j)-C_1)^{-1}\nabla\Phi[\hat\pi_n](\hat\nu^*_j) \rangle\cr
\end{split}
\end{equation}
{\em Non-degeneracy assumption 1)} implies together with the twice continuous differentiability of $F$ 
that there exists a positive constant $\tilde K$ such that
\begin{equation}
 \langle x,H\Phi[\xi](\hat\nu^*_j)x\rangle\geq \tilde K ||x||^2
\end{equation}
for all $\xi$ in a neighborhood of $\pi$. 
Noticing that $||\nabla\Phi[\hat\pi_n](\hat\nu^*_j)||\leq c ||\hat\pi_n-\pi||$ we have 
\begin{equation}
 -\frac{K||\hat\pi_n-\pi||^2}{2}\leq\displaystyle\inf_{\hat\nu\in B(\e,\pi\hat\nu_j^*)}(\Phi[\hat\pi_n](\hat\nu)-\Phi[\hat\pi_n](\hat\nu_j^*))\leq 0
\end{equation} 
with $K=\frac{c^2}{\tilde K-C_1}$ which is positive for $\e_0$ sufficiently small. 
$\Cox$
\vskip .1in
\noindent
From the last right-hand side of \eqref{rho2} we have 
 \begin{equation}
 \begin{split}\label{const_C}
&\frac{1}{1+\displaystyle\sum_{j\neq i}^{k}\frac{\displaystyle\sum_{\hat \nu:\hat\pi_n \hat \nu \in B(\e,\pi\hat\nu^*_j)}e^{-n(\Phi[\hat\pi_n](\hat\nu)-\Phi[\hat\pi_n](\hat\nu_j^*))}e^{-n\Phi[\hat\pi_n](\hat\nu_j^*)}}{\displaystyle\sum_{\hat \nu:\hat\pi_n \hat \nu \in B(\e,\pi\hat\nu^*_i)}e^{-n(\Phi[\hat\pi_n](\hat\nu)-\Phi[\hat\pi_n](\hat\nu_i^*))}e^{-n\Phi[\hat\pi_n](\hat\nu_i^*)}\prod_{b\in E'}(\mid \Lambda_n(b)\mid +1)^{-\mid E'\mid}}}\cr
\geq&\frac{1}{1+\displaystyle\sum_{j\neq i}^{k}\frac{\displaystyle\sum_{\hat \nu:\hat\pi_n \hat \nu \in B(\e,\pi\hat\nu^*_j)}e^{-n(\displaystyle\inf_{{\hat\nu\in}\atop{ \hat M_n\cap B(\e,\pi\hat\nu_j^*)}}(\Phi[\hat\pi_n](\hat\nu)-\Phi[\hat\pi_n](\hat\nu_j^*)))}e^{-n\Phi[\hat\pi_n](\hat\nu_j^*)}}{e^{-n(\Phi[\hat\pi_n](\tilde\nu)-\Phi[\hat\pi_n](\hat\nu_1^*)))}e^{-n\Phi[\hat\pi_n](\hat\nu_i^*)}\prod_{b\in E'}(\mid \Lambda_n(b)\mid +1)^{-\mid E'\mid}}}\cr
\geq&\frac{1}{1+\displaystyle\sum_{j\neq i}^{k}\frac{\displaystyle\sum_{\hat \nu:\hat\pi_n \hat \nu \in B(\e,\pi\hat\nu^*_j)}
e^{\frac{K n \Vert \hat\pi_n-\pi\Vert^2}{2}} e^{-n\Phi[\hat\pi_n](\hat\nu_j^*)}}{e^{-C} e^{-n\Phi[\hat\pi_n](\hat\nu_i^*)}\prod_{b\in E'}(\mid \Lambda_n(b)\mid +1)^{-\mid E'\mid}}}\cr
 \end{split}
\end{equation}
In the first inequality we have chosen $\tilde \nu$ as a best-approximation of $\hat \nu_i^*$ in $\hat M_n$ 
to get rid of the sum in the denominator of the denominator. 
In the second inequality we have used Prop.\eqref{pp2.5}, and moreover the bound on the corresponding discretization error of the order $1/n$ and the 
 uniform boundedness of the first derivative of $\Phi$. The sums over measures 
 in balls only give rise to polynomial constants which are swallowed by the terms 
 in the exponential (as we will see, because the random terms lifting 
 the degeneracy between the minimizers will be of order squareroot.)

Now to the lowest order in $\hat\pi_n-\pi$, we have
\begin{equation}
\begin{split}\label{taylor}
&\Phi[\hat\pi_n](\hat\nu^*_i)=\Phi[\pi](\hat\nu^*_i)+\Phi_{\pi}[\pi](\hat\nu^*_i)(\hat\pi_n-\pi)+o(\parallel\hat\pi_n-\pi\parallel)\cr
\end{split}
\end{equation}
So the last right-hand side of \eqref{const_C} becomes
\begin{equation}
\begin{split}\label{lemmarho2}
\geq&\frac{1}{1+\displaystyle\sum_{j\neq i}^{k} e^{\frac{ K n \Vert \hat\pi_n-\pi\Vert^2}{2}+C}e^{-n\langle B_{\hat\nu^*_i}-B_{\hat\nu^*_j},\hat\pi_n-\pi\rangle}e^{-n\cdot o(\parallel\hat\pi_n-\pi\parallel)}\prod_{b\in E'}(\mid \Lambda_n(b)\mid+1)^{2\mid E\mid}}\cr
\end{split}
\end{equation}
We are considering $n$ sufficiently large such that there is at least one element in \\
$\{\hat \nu:\hat\pi_n \hat \nu \in B(\e,\pi\hat\nu^*_i)\}$.\\
For $\eta \in \HH_{i,n,0}^{\t}$ we have that 
\begin{equation}
 \begin{split}\label{rhoepsilon}
&\r^{\e}[\eta](n)(\pi\hat\nu^*_i) > 1-r(n)\cr
\end{split}
\end{equation}
Indeed we defined the {\em good-set} $\HH_{i,n,0}^{\t}$ in such a way that 
$n \Vert \hat\pi_n-\pi\Vert^2\leq n^{\frac{\t}{2}}$ and $n \langle B_{\hat\nu^*_i}-B_{\hat\nu^*_j},\hat\pi_n-\pi\rangle\geq n^{\t}$. 
Here we see the reason for the choice of the cutoff.
%


In order to prove the Lemma\eqref{concentrationH} let us write 
\begin{equation}
 \begin{split}\label{mu}
&\mu_{F,n}[\eta(1),\dots,\eta(n)]( L_n\in B(\e,\pi \hat\nu^*_i))\cr
=&\r^{\e}[\eta](n)(\pi\hat\nu^*_i) (1-\mu_{F,n}[\eta(1),\dots,\eta(n)](d( L_n,\pi M^*)\geq \e))\cr
 \end{split}
\end{equation}
Now we can use the concentration property for the empirical distribution saying that 
$\forall\e>0$ and for all $\eta\in \HH_{i,n,0}^{\t}$ we have 
\begin{equation}\label{concentrantion-prop}
  \mu_{F,n}[\eta(1),\dots,\eta(n)](d( L_n,\pi M^*)\geq \e)\leq \hat r(n,\e)
\end{equation}
with $\lim_{n\uparrow \infty}\hat r(n,\e)=0$ for all positive $\e$. 
 
This concentration property is a consequence of the bound 

\begin{equation}
\begin{split}\label{1.3.2}
&\mu_{F,n}[\eta(1),\dots,\eta(n)](d( L_n,\pi M^*) \geq \e)\cr
& \leq \prod_{b\in E'} (n\hat \pi_n(b)+1)^{2 |E|}
\exp\left(- n \inf_{
{\hat \nu \in \hat M_{n}:}\atop
{d(\hat\pi_n\hat \nu, \pi M^*) \geq \e }} 
\Phi[\hat \pi_n](\hat \nu)+n \inf_{
\hat \nu' \in \hat M_{n}}\Phi[\hat \pi_n](\hat \nu')
\right)\cr 
& \leq  \prod_{b\in E'} (n\hat \pi_n(b)+1)^{2 |E|}
e^{K_2 n  \Vert \hat\pi_n-\pi\Vert + C_2}
\exp\left(- n \inf_{
{\hat \nu \in \PP(E)^{|E'|}:}\atop
{d(\pi \hat \nu,\pi M^*) \geq \e }} 
\Phi[\pi](\hat \nu)+n \inf_{
\hat \nu' \in \PP(E)^{|E'|}}\Phi[\pi](\hat \nu')
\right)\cr 
\end{split} 
\end{equation}
where in the second inequality we have used the Lipschitz property of 
$\Phi$ w.r.t. $\pi$ and the control of the discretization error. On the {\em good-sets} 
we have $n  \Vert \hat\pi_n-\pi\Vert \leq n^{\frac{1}{2}+\frac{\t}{4}}$, while 
the quadratic nature of the minima gives us a term of exponential decay in $n$ from 
the rightmost exponential,  for any fixed $\e>0$. This proves the concentration property.
So the Lemma follows from \eqref{rhoepsilon},\eqref{mu} and \eqref{concentrantion-prop}. $\Cox$
\bigskip 


%


%

%


Having proved, for a particular choice of the disorder variables, the concentration of the empirical distribution around a given minimizer, the following Lemma represents the natural extension to averages.
\begin{lem}\label{conti}
For any real-valued continuous function $g$ on $\PP(E)$ the following holds:
\begin{equation}
\begin{split}\label{cong}
&|\r[\eta](n)(g)-g(\pi \hat \nu^*_j)|\leq  \tilde{r}(n),\;\;\;\;\forall \eta\in \HH_{j,n,0}^{\t} 
\end{split} 
\end{equation}
where $\lim_{n\uparrow\infty}\tilde{r}(n)=0$. 
\end{lem}
\vskip .1in
\noindent
\textbf{Proof: } Let $B(\e,\pi\hat{\nu}^*_j)$ be an $\e$-ball around the measure $\pi\hat{\nu}^*_j$. Then for any $\e>0$ and integer $n$,
\begin{equation*}\begin{split}
&|\r[\eta](n)(g)-g(\pi \hat \nu^*_j)|=\bigg|\r[\eta](n)\Big(1_{B(\e,\pi\hat{\nu}^*_j)}(g-g(\pi\hat{\nu}^*_j))\Big)+\r[\eta](n)\Big(1_{B^c(\e,\pi\hat{\nu}^*_j)}(g-g(\pi\hat{\nu}^*_j))\Big)\bigg|\cr
&\leq \sup_{\nu\in B(\e,\pi\hat\nu^*_j)}|g(\nu)-g(\pi\hat\nu^*_j)|
+ 2 \Vert g \Vert_{\infty} 
\;\r[\eta](n)\Big(B^c(\e,\pi\hat\nu^*_j)\Big)\cr
\end{split}
\end{equation*} 
holds.
Choosing first $\e$ sufficiently small and then $n$ sufficiently large proves the lemma. 
$\Cox$
\bigskip 

Now comes the study of how the probability of the {\em good-sets} $\HH_{j,n,l}^{\t}$ behaves in the limit $n\uparrow\infty$. Out of this analysis the weights \eqref{nondeg4} will arise.
The fundamental step is that the limit will not depend on any  finite number $l$ of coordinates $\eta$, while the corresponding tail will provide, using CLT, the longed weights. This together with the Stone-Weierstrass theorem and the Lemma\eqref{conti} are the overriding tools for proving Theorem\eqref{thm2}. \\
Let us start the analysis looking at the $n,l$-dependent {\em good-sets}  $\HH_{i,n,l}^{\t}$ in a slightly different way.\\
For any $l<n$, we have
\begin{equation}
 \begin{split}\label{partitionLofX}
\tiny&\HH_{i,n,0}^{\t} =\left\lbrace \eta \in (E')^n: \frac{\sqrt{nl}}{n}X_{[1,l]}[\eta]+\frac{\sqrt{n(n-l)}}{n}X_{[l+1,n]}[\eta]\in R_{i,n}^{\t}\right\rbrace \cr 
\end{split}
\end{equation}

Saying that $X_{[1,n]}[\eta]\in R_{i,n}^{\t}$ it means
\begin{equation}
\begin{split}\label{lemmaball}
\tiny &a_n\langle X_{[1,l]}[\eta],  B_i \rangle+ b_n\langle X_{[l+1,n]}[\eta],  B_i \rangle-\max_{k\neq i} \left( a_n\langle X_{[1,l]}[\eta],  B_k \rangle+b_n\langle X_{[l+1,n]}[\eta],  B_k \rangle\right)>{\delta_n}\cr
\end{split} 
\end{equation} 
and $\Vert X_{[1,n]}[\eta]\Vert\leq n^{\frac{\t}{4}}$, 
where $a_n=\frac{\sqrt{nl}}{n}$ and $b_n=\frac{\sqrt{n(n-l)}}{n}$\\

Now let us define a subregion of $\HH_{i,n,0}^{\t}$, namely $\HH_{i,n,0}^{\t}(l)$ as follows
\begin{equation}
\begin{split}\label{nondeg5}
&\HH_{i,n,0}^{\t}(l):=
\biggl\{ \eta \in (E')^n:\cr & a_n\langle X_{[1,l]}[\eta],  B_i \rangle+b_n\langle X_{[l+1,n]}[\eta],  B_i \rangle-\max_{k\neq i} \left( a_n\langle X_{[1,l]}[\eta],  B_k \rangle\right) -\max_{k\neq i}\left( b_n\langle X_{[l+1,n]}[\eta],  B_k \rangle\right)>{\delta_n}, \cr & \text{ and } \Vert X_{[1,n]}[\eta]\Vert\leq n^{\frac{\t}{4}}\biggr\}  \cr 
\end{split} 
\end{equation} 
{ \bf Remark:} While $\HH_{i,n,0}^{\t}$ does not depend on $l$ , $\HH_{i,n,1}^{\t}(l)$ does,
indeed the partitioning might change the {\em max-value}.
\bigskip \\

It is worthwhile mentioning the following results.
\begin{lem}\label{probforH}
For any integer $l$, $\P(\HH_{i,n,0}^{\t} \setminus \HH_{i,n,0}^{\t}(l))$ goes to zero in the limit $n\uparrow \infty$.
\end{lem}

\noindent
\vskip .1in
\noindent
\textbf{Proof:}  Note that 
\begin{equation}
\begin{split}\label{nondegh1}
\HH_{i,n,0}^{\t} \setminus \HH_{i,n,0}^{\t}(l)\subseteq 
\biggl\{ &\eta \in(E')^n:\max_{k\neq i} \langle a_nX_{[1,l]}[\eta],  B_k \rangle +\max_{k\neq i} \langle b_nX_{[l+1,n]}[\eta],  B_k \rangle+{\delta_n} \cr
&\geq\langle a_nX_{[1,l]}[\eta]+b_nX_{[l+1,n]}[\eta],  B_i \rangle 
>\max_{k\neq i}\langle a_nX_{[1,l]}[\eta]+b_nX_{[l+1,n]}[\eta],  B_k \rangle+{\delta_n} 
\biggr\}.\cr 
\end{split}
\end{equation}
Now
\begin{equation}
 \begin{split}\label{inclusion11}
\HH_{i,n,0}^{\t} &\setminus \HH_{i,n,0}^{\t}(l)
\subseteq \biggl\{ \eta:\frac{C(l)}{\sqrt{n}}+\max_{k\neq i} \langle b_nX_{[l+1,n]}[\eta],  B_k \rangle+{\delta_n} \cr
&\geq\langle a_nX_{[1,l]}[\eta]+b_nX_{[l+1,n]}[\eta],  B_i \rangle 
> - \frac{C(l)}{\sqrt{n}}+\max_{k\neq i} \langle b_nX_{[l+1,n]}[\eta],  B_k \rangle+{\delta_n}\biggr\}\cr 
\end{split}
\end{equation}
where $C(l)=\sqrt{l}\max_\eta\max_k\mid\langle X_{[1,l]}[\eta],  B_k \rangle\mid\leq \sqrt{l}\max_k\Vert B_k\Vert_{\infty}$.\\
The set on the right-hand side of \eqref{inclusion11} can be written as
\begin{equation}
 \begin{split}\label{intersection}
&\biggl\{ \eta:\frac{C(l)}{\sqrt{n}}+\max_{k\neq i} \langle b_nX_{[l+1,n]}[\eta],  B_k \rangle+{\delta_n} 
\geq a_n\langle X_{[1,l]}[\eta],B_i\rangle+b_n\langle X_{[l+1,n]}[\eta],  B_i \rangle\biggr\}\cr
&\cap\biggl\{\eta:- \frac{C(l)}{\sqrt{n}}+\max_{k\neq i} \langle b_nX_{[l+1,n]}[\eta],  B_k \rangle+{\delta_n}< a_n\langle X_{[1,l]}[\eta],B_i\rangle+b_n\langle X_{[l+1,n]}[\eta],  B_i \rangle\biggr\}\crcr 
&=\biggl\{ \eta: (\frac{C(l)}{\sqrt{n}}-a_n\langle X_{[1,l]}[\eta],B_i\rangle+{\delta_n})b_n^{-1}
\geq \langle X_{[l+1,n]}[\eta],  B_i \rangle-\max_{k\neq i} \langle X_{[l+1,n]}[\eta],  B_k \rangle \biggr\}\cr
&\cap\biggl\{ \eta: (-\frac{C(l)}{\sqrt{n}}-a_n\langle X_{[1,l]}[\eta],B_i\rangle+{\delta_n})b_n^{-1}
< \langle X_{[l+1,n]}[\eta],  B_i \rangle-\max_{k\neq i} \langle X_{[l+1,n]}[\eta],  B_k \rangle \biggr\}\cr
&\subset\biggl\{ \eta: (\frac{2C(l)}{\sqrt{n}}+{\delta_n})b_n^{-1}
\geq \langle X_{[l+1,n]}[\eta],  B_i \rangle-\max_{k\neq i} \langle X_{[l+1,n]}[\eta],  B_k \rangle \biggr\}\cr
&\cap\biggl\{ \eta: (-\frac{2C(l)}{\sqrt{n}}+{\delta_n})b_n^{-1}
< \langle X_{[l+1,n]}[\eta],  B_i \rangle-\max_{k\neq i} \langle X_{[l+1,n]}[\eta],  B_k \rangle \biggr\}\cr
\end{split}
\end{equation}
Let's define 
\begin{equation}
\begin{split}\label{phi}
 &\phi^i(X_{[l+1,n]}[\eta]):=\langle X_{[l+1,n]}[\eta],  B_i \rangle-\max_{k\neq i} \langle X_{[l+1,n]}[\eta],  B_k \rangle\cr
\end{split}
\end{equation}
\\
So we have 
\begin{equation}
 \begin{split}
  \P(\HH_{i,n,0}^{\t} \setminus \HH_{i,n,0}^{\t}(l))\leq\P\left( \biggl\{\eta:\phi^i(X_{[l+1,n]}[\eta])\in b_n^{-1}\left( {\delta_n}-\frac{2C(l)}{\sqrt{n}},{\delta_n}+\frac{2C(l)}{\sqrt{n}}\right) \biggr\}\right) \cr
 \end{split}
\end{equation}

To take care of the $n$-dependence of the interval it's enough to notice that, $\forall\e>0\;\exists\; \bar n(\e)$ such that, for all $n>\bar n(\e)$ the following holds
\begin{equation}
 \begin{split}
b_n^{-1} \left( {\delta_n}-\frac{2C(l)}{\sqrt{n}},{\delta_n}+\frac{2C(l)}{\sqrt{n}}\right)\subset(-\e,\e)\cr
 \end{split}
\end{equation}

So
\begin{equation}
 \begin{split}
  & \lim_{n\uparrow\infty}\P\left( \biggl\{\eta:\phi^i(X_{[l+1,n]}[\eta])\in b_n^{-1}\left( {\delta_n}-\frac{2C(l)}{\sqrt{n}},{\delta_n}+\frac{2C(l)}{\sqrt{n}}\right) \biggr\}\right)\cr &\leq  \lim_{n\uparrow\infty}\P\left( \biggl\{\eta:\phi^i(X_{[l+1,n]}[\eta])\in (-\e,\e) \biggr\}\right) \cr
 \end{split}
\end{equation}
By the multidimensional CLT we have
\begin{equation}
 \begin{split}
& \lim_{n\uparrow\infty}\P\left( \biggl\{\eta:\phi^i(X_{[l+1,n]}[\eta])\in (-\e,\e)\biggr\}\right)=\P_{\pi}\left( \phi^i(G)\in (-\e,\e)\right)  \cr
 \end{split}
\end{equation}
where $G$ is a centered Gaussian variable.\\
Taking the limit $\e\downarrow 0$ and using the {\em non-degeneracy assumption 2)}, the lemma is proved.
$\Cox$
\bigskip 
We have just seen that, for any fixed integer $l$, there is a subregion of the {\em good-set} which will not play any role in the limit $n \uparrow\infty$. We focus now on the probability of the main part of the {\em good-set}, especially on how its limit does not depend on any finite number of $\eta$-coordinates. Let us formalize the previous heuristic.\\
The condition \eqref{nondeg5} defining $\HH_{i,n,0}^{\t}(l)$ can also be written as
\begin{equation}
\begin{split}\label{nondeg3}
&\langle X_{[l+1,n]}[\eta],  B_i \rangle-\max_{k\neq i}\langle X_{[l+1,n]}[\eta],  B_k \rangle>\frac{a_n}{b_n}\left( \max_{k\neq i}\langle X_{[1,l]}[\eta],  B_k \rangle- \langle X_{[1,l]}[\eta],  B_i \rangle\right) + b_n^{-1}{\delta_n}\cr & \text{ and }\Vert X_{[1,n]}[\eta]\Vert\leq n^{\frac{\t}{4}}. 
\end{split}
\end{equation}
 Define the following sets
\begin{equation}
 \begin{split}
 \tiny & A_{i,n}^{{\t}}(l):=\bigg\{\eta:\langle  X_{[l+1,n]}[\eta],  B_i \rangle-\max_{k\neq i} \langle
 \tiny X_{[l+1,n]}[\eta],  B_k \rangle> -\frac{a_n}{b_n}C_2(l)+b_n^{-1} {\delta_n}, \Vert X_{[l+1,n]}[\eta]\Vert\leq b_n ^{-1}n^{\frac{\t}{4}}\bigg\}\cr
\tiny & B_{i,n}^{{\t}}(l):=\bigg\{\eta:\langle X_{[l+1,n]}[\eta],  B_i \rangle-\max_{k\neq i} \langle 
\tiny X_{[l+1,n]}[\eta],  B_k \rangle> \frac{a_n}{b_n}C_2(l)+b_n^{-1} {\delta_n},\Vert X_{[l+1,n]}[\eta]\Vert\leq b_n ^{-1}(n^{\frac{\t}{4}}-\tilde C_2(l))\bigg\}\cr
 \end{split}
\end{equation}
where $C_2(l)=\max_{\eta}\mid\max_{k\neq i} \langle  
\tiny X_{[1,l]}[\eta],  B_k \rangle- \langle X_{[1,l]}[\eta],  B_i \rangle\mid\leq 2 \max_k \Vert B_k\Vert_{\infty}$, and $\tilde C_2(l)=\max_{\eta}\Vert X_{[1,l]}[\eta]\Vert$.
These maxima give us the intended independence of the set from $\eta\in (E')^l$ and we have\\
\begin{equation}
 \begin{split}\label{partition}
  &B_{i,n}^{{\t}}(l)=(E')^l\times\HH_{i,n,l}^{1,\t}\cr
&A_{i,n}^{{\t}}(l)=(E')^l\times\HH_{i,n,l}^{2,\t}\cr
 \end{split}
\end{equation}

 where
\begin{equation}
\begin{split}\label{region(n-L)}
  &\HH_{i,n,l}^{1,\t}=\biggl\{\eta\in (E')^{n-l}:\phi^i(X_{[l+1,n]}[\eta])>b_n^{-1}\left( {\delta_n} + a_n C_2(l)\right),\text{ and } \Vert X_{[l+1,n]}[\eta]\Vert\leq b_n ^{-1}(n^{\frac{\t}{4}}-\tilde C_2(l))  \bigg\}\cr
&\HH_{i,n,l}^{2,\t}=\biggl\{\eta\in (E')^{n-l}:\phi^i(X_{[l+1,n]}[\eta])>b_n^{-1}\left( {\delta_n} - a_n C_2(l)\right),\text{ and }\Vert X_{[l+1,n]}[\eta]\Vert\leq b_n ^{-1}n^{\frac{\t}{4}}\bigg\}\cr
\end{split}
\end{equation}

The following holds
\begin{equation}
 \begin{split}\label{inclusion}
 & A_{i,n}^{{\t}}(l)\supseteq \HH_{i,n,1}^{\t}(l) \supseteq B_{i,n}^{{\t}}(l)\cr
&\HH_{i,n,l}^{1,\t}\subseteq\HH_{i,n,l}^{2,\t}\cr
 \end{split}
\end{equation}

\begin{lem}
For any integer $l$, $\P(\HH_{i,n,l}^{2,\t} \setminus \HH_{i,n,l}^{1,\t})$ goes to zero in the limit $n\uparrow \infty$.
\end{lem}
\noindent
\vskip .1in
\noindent
\textbf{Proof:} 
\begin{equation}
\begin{split}\label{nonso}
& \HH_{i,n,l}^{2,\t} \setminus \HH_{i,n,l}^{1,\t}\subseteq 
\biggl\{\eta:\frac{a_n}{b_n}C_2(l)+b_n^{-1}{\delta_n}\geq\cr
& \langle X_{[l+1,n]}[\eta],  B_i \rangle-\max_{k\neq i} \langle X_{[l+1,n]}[\eta],  B_k \rangle>-\frac{a_n}{b_n}C_2(l)+b_n^{-1}{\delta_n}
\biggr\}.=\cr
& \biggl\{\eta: b_n^{-1}\left( {\delta_n} + a_n C_2(l)\right) \geq \phi^i(X_{[l+1,n]}[\eta])>b_n^{-1}\left( {\delta_n} - a_n C_2(l)\right) \biggr\}\cr 
\end{split}
\end{equation}
By the same argument we have used in Lemma\eqref{probforH}, we have \\ $\P(\HH_{i,n,l}^{2,\t} \setminus \HH_{i,n,l}^{1,\t})\longrightarrow 0$ in the limit $n\uparrow\infty$.
$\Cox$
\bigskip 

\begin{lem}\label{pesi}
 For any integer $l$, $ \lim_{n\uparrow\infty}\P(\HH_{i,n,l}^{1,\t})=\P_{\pi}(G\in R_{i})$
where $G\sim\NN(0,\Sigma)$.
\end{lem}
\textbf{Proof:} From the previous lemma we know that
\begin{equation}
 \begin{split}\label{raletion}
  & \lim_{n\uparrow\infty}\P(\HH_{i,n,1}^{\t}(l))=\lim_{n\uparrow\infty}\P(B_{i,n}^{{\t}}(l))\cr
\text{and}\cr 
&\lim_{n\uparrow\infty}\P(\HH_{i,n,l}^{2,\t})=\lim_{n\uparrow\infty}\P(\HH_{i,n,l}^{1,\t})\cr
\end{split}
\end{equation}
Now $\forall\e>0\;\exists\; n_0(\e)$ such that for all $n>n_0(\e)$ the following holds
\begin{equation}
 \gamma_{-\e}^n\supset\HH_{i,n,l}^{1,\t}\supset\gamma_{\e}^n
\end{equation}
where $\gamma_{\e}^n= \{ \eta:\phi^i(X_{[l+1,n]}[\eta])>\e \} $.
Therefore
\begin{equation}
 \lim_{n\uparrow\infty}\P(\gamma_{-\e}^n)\geq\lim_{n\uparrow\infty}\P(\HH_{i,n,l}^{1,\t})\geq\lim_{n\uparrow\infty}\P(\gamma_{\e}^n)
\end{equation}
Applying the CLT to both the right and the left-hand side\\
and taking the limit for $\e\downarrow0$ we have
\begin{equation}\label{gaussianweight}
 \lim_{n\uparrow\infty}\P(\HH_{i,n,l}^{1,\t})=\P_{\pi}(G\in R_{i})
\end{equation}
where $G\sim\NN(0,\Sigma)$.

$\Cox$
\noindent
\vskip .1in
\noindent
Let us now summarize what we have done above for the decompositions of the various regions of the $\eta$-configuration space.
\begin{equation}
\begin{split}\label{nondegh5}
&1_{\HH_{i,n,0}^{\t}}=1_{\HH_{i,n,0}^{\t}(l)}+1_{\HH_{i,n,0}^{\t} \setminus \HH_{i,n,0}^{\t}(l)}\cr
&1_{B_{i,n}^{\t}(l)}=1_{(E')^l}1_{\HH_{i,n,l}^{1,\t}}\cr
&1_{A_{i,n}^{\t}(l)}=1_{(E')^l}1_{\HH_{i,n,l}^{2,\t}}
\end{split}
\end{equation}
To state our next result let us fix one more notation. \\
We let $\Psi$ be  a continuous  real-valued function on $\PP(\PP(E))\times (E')^m$, for some  positive integer $m$. 

\begin{lem}\label{limH_n(L)}
Suppose $\Psi$ is as above. Then under the {\em non-degeneracy assumptions} 1) and 2) the following holds:
\begin{equation}
\lim_{n\uparrow\infty}\int_{\HH_{i,n,0}^{\t}}\P_\pi(d\eta)\Psi(\rho[\eta](n),\eta)=w_i\int_{(E')^m} \pi^{\otimes m}(d\eta)  \Psi(\d_{\pi\hat\nu_i},\eta), 
\end{equation}
where $\pi^{\otimes m}(d\eta)=\prod_{k=1}^m\pi(d\eta_k)$, $w_i=\P(G\in R_i)$ with $G\sim\NN(0,\Sigma)$
\end{lem}
\noindent
\textbf{Proof: }
Set $l=m$,
\begin{equation}
 \int_{\HH_{i,n,0}^{\t}}\P_\pi(d\eta)\Psi(\rho[\eta](n),\eta)=\int_{\HH_{i,n,0}^{\t}}\P_\pi(d\eta)(\Psi(\rho[\eta](n),\eta)-\Psi(\d_{\pi\hat\nu_i},\eta))+\int_{\HH_{i,n,0}^{\t}}\P_\pi(d\eta)\Psi(\d_{\pi\hat\nu_i},\eta)
\end{equation}
We can assume that $\Psi$ is of the form 
$\Psi(\r,\eta)=\tilde \Psi(\r(g_1),\dots, \r(g_l),\eta_{[1,m]})$ for a finite $l$ with continuous 
and bounded $g_i$'s, and continuous $ \tilde \Psi$. 
So, together with the Lemma\eqref{conti} we have that the first term in the left-hand side is going to $0$ in the limit $n\uparrow\infty$.\\
Now from the first equality of \eqref{nondegh5} 
\begin{equation*}\begin{split}\label{ihdn}
\int_{\HH_{i,n,0}^{\t}}\P_\pi(d\eta) \Psi(\d_{\pi\hat\nu_i},\eta) 
&=\int_{\HH_{i,n,0}^{\t}(l)}\P_\pi(d\eta) \Psi(\d_{\pi\hat\nu_i},\eta) +\int_{\HH_{i,n,0}^{\t}\setminus \HH_{j,n,1}^{\t}(l)} \P_\pi(d\eta) \Psi(\d_{\pi\hat\nu_i},\eta). 
%
\end{split} 
\end{equation*}
Under the {\em non-degeneracy assumption 2)} the second term on the right-hand side of the above equation  plays no role in the limit, indeed
\begin{equation}
 \int_{\HH_{i,n,0}^{\t}\setminus \HH_{i,n,0}^{\t}(l)} \P_\pi(d\eta) \Psi(\d_{\pi\hat\nu_i},\eta)\leq\parallel \Psi\parallel_{\infty}\P(\HH_{i,n,0}^{\t}\setminus \HH_{i,n,0}^{\t}(l))
\end{equation}
and from the Lemma\eqref{probforH} $\P(\HH_{i,n,0}^{\t} \setminus \HH_{i,n,0}^{\t}(l))$ goes to zero in the limit $n\uparrow \infty$.
\vskip .1in
\noindent
Observe from the first inclusion relation of \eqref{inclusion} that
\begin{equation}
 \begin{split}\label{integralinequolities}
  &\int_{B_{i,n}^{\t}(l)}\P_\pi(d\eta) \Psi(\d_{\pi\hat\nu_i},\eta)\leq\int_{\HH_{i,n,0}^{\t}(l)}\P_\pi(d\eta) \Psi(\d_{\pi\hat\nu_i},\eta)\leq\int_{A_{i,n}^{\t}(l)}\P_\pi(d\eta) \Psi(\d_{\pi\hat\nu_i},\eta)\cr
 \end{split}
\end{equation}

Next, observe  from  \eqref{partition} that
\begin{equation}
 \begin{split}\label{integralinequolitiespartitioned}
 \tiny &\int_{(E')^l} \pi^{\otimes l}(d\eta) \Psi(\d_{\pi\hat\nu_i},\eta)\int_{\HH_{i,n,l}^{1,\t}}\pi^{\otimes n-l}(d\eta)\leq\int_{\HH_{i,n,0}^{\t}(l)}\P_\pi(d\eta) \Psi(\d_{\pi\hat\nu_i},\eta)\cr
&\int_{\HH_{i,n,0}^{\t}(l)}\P_\pi(d\eta) \Psi(\d_{\pi\hat\nu_i},\eta)\leq \int_{(E')^l} \pi^{\otimes l}(d\eta) \Psi(\d_{\pi\hat\nu_i},\eta)\int_{\HH_{i,n,l}^{2,\t}}\pi^{\otimes n-l}(d\eta) \cr
 \end{split}
\end{equation}

Taking the limit $n\uparrow\infty$ we obtain
\begin{equation}
 \begin{split}\label{limitH(L)}
\lim_{n\uparrow\infty}\int_{\HH_{i,n,0}^{\t}(l)}\P_\pi(d\eta) \Psi(\d_{\pi\hat\nu_i},\eta)=
  \int_{(E')^l} \pi^{\otimes l}(d\eta) \Psi(\d_{\pi\hat\nu_i},\eta)\lim_{n\uparrow\infty}\int_{\HH_{i,n,l}^{2,\t}}\pi^{\otimes n-l}(d\eta) 
 \end{split}
\end{equation}
and using Lemma\eqref{pesi} we are done.

$\Cox$
\bigskip

Now we have provided all the ingredients, and so the proof of the Theorem \eqref{thm2} is straightforward.
\begin{equation}
\begin{split}\label{nondeg3}
&\int\P_\pi(d\eta)\Psi(\r[\eta](n),\eta)\cr
&=\sum_{i=1}^k \int\P_\pi(d\eta)\Psi(\r[\eta](n),\eta)1_{\HH_{i,n,1}^{\t}}(\eta)+ \int\P_\pi(d\eta)\Psi(\r[\eta](n),\eta)1_{(\HH_{n,1}^{\t})^c}(\eta)
\cr
\end{split} 
\end{equation}
Clearly for bounded $\Psi$ one has 
\begin{equation}
\begin{split}\label{nondeg3}
&\Bigl | \int\P_\pi(d\eta)\Psi(\r[\eta](n),\eta)1_{(\HH_{n,0}^{\t})^c}(\eta)\Bigr | 
\leq \Vert \Psi \Vert_{\infty} \P_\pi((\HH_{n,0}^{\t})^c(\eta))
\cr
\end{split} 
\end{equation}
and the {\em non-degeneracy assumption 2)} tells us that this term will not play any role in the limit $n\uparrow\infty$.
For every summand of the first term, by Lemma\eqref{limH_n(L)} we have
\begin{equation}
\lim_{n\uparrow\infty}\int_{\HH_{i,n,0}^{\t}}\P_\pi(d\eta)\Psi(\rho[\eta](n),\eta)=w_i\int_{(E')^m} \pi^{\otimes m}(d\eta)  \Psi(\d_{\pi\hat\nu_i},\eta), 
\end{equation}
where $w_i=\P(G\in R_i)$ with $G\sim\NN(0,\Sigma)$.\\
Therefore
\begin{equation}
\begin{split}
&\lim_{n\uparrow\infty}\int\P_\pi(d\eta)\Psi(\r[\eta](n),\eta)=\sum_{i=1}^k\int\P_\pi(d\eta)\Psi(\r,\eta)w_i\d_{\d_{\pi \hat \nu_i}}(d\r)
\end{split} 
\end{equation}
Looking now at the definition of the AW-metastate, we can identify the joint distribution $K$ we are interested in
\begin{equation}
\begin{split}
&K(d\r,d\eta)=\sum_{i=1}^k\P_\pi(d\eta)w_i\d_{\d_{\pi \hat \nu_i}}(d\r)\cr 
&\Longrightarrow K(d\r|\eta)=\sum_{i=1}^kw_i\d_{\d_{\pi \hat \nu_i}}(d\r)
\end{split} 
\end{equation}

 $\Cox$
\bigskip
\bigskip 

\section{The Metastate on the level of states}

Let us go from the global perspective (talking about the empirical mean) to the local view (talking about finitely many variables $\s_1,\dots,\s_k$). In different words, we are fixing a subpopulation of finite size, and we are asking how it will behave when we couple it to a large system whose size $n$ will be let tend to infinity.
Let us introduce a metric on the space of probability measures $\mu,\mu'\in \PP(E^{\infty})$ by
\begin{equation}\label{metric}
 d(\mu,\mu')=\displaystyle\sum_{i=1}^{\infty}2^{-i}||\mu-\mu'||_i
\end{equation}
where
\begin{equation}
||\mu-\mu'||_i:=\frac{1}{2}\displaystyle\sum_{\o_1,\dots,\o_i}|\mu(\o_1,\dots,\o_i)-\mu'(\o_1,\dots,\o_i)|
\end{equation}
is the total variation norm of the restriction of the measure to the first $i$ coordinates.\\
The statement about the metastate promised in the main theorem implies in particular that, for all $\e>0$  
\begin{equation}\label{limequation}
 \lim_{n\uparrow\infty}\P( d(\mu_{F,n}[\eta],ext(\GG[\eta]))>\e)=0
\end{equation}
where
\begin{equation}\label{gibbssimplex}
 \GG[\eta]=\{\displaystyle\sum_{\hat\nu\in M^*}p_{\hat\nu}\mu_{\hat\nu}[\eta],\;p_{\hat\nu}\in \PP(\pi M^*)\}
\end{equation}
and $\mu_{\hat\nu}[\eta](\cdot)=\displaystyle\prod_{i=1}^{\infty}\gamma[\eta(i)](\cdot|\pi\hat\nu)$.
Thoughout this chapter we identify $\mu_{F,n}[\eta]$ with the infinite-volume measure which 
is obtained by tensorization with the equidistribution for sites outside of $\{1,\dots,n\}$. 

We will in fact prove that
\begin{equation}\label{uniform}
\begin{split}
 &\lim_{n\uparrow\infty}
 \sup_{\eta \in \HH_{i,n,0}^\t } d(\mu_{F,n}[\eta],\mu_{\hat\nu_i^*}[\eta])=0
\end{split}
\end{equation}
where $\HH_{i,n,0}^\t$ are the disorder sets ensuring the dominance of the $i$-th minimizer. 
Let us remark that it can not be expected in general that $ \lim_{n\uparrow\infty}d(\mu_{F,n}[\eta],ext(\GG[\eta])=0$ for $\P$-a.e. $\eta$ holds, as already the example of the random field Ising model discussed 
in \cite{Ku97} shows, due to the empirical distribution $\hat \pi_n$ passing regions of "ties" outside 
of the good sets infinitely often. 
\\

So we are about to prove that the possible limiting distributions will be product measures of a particular sort.
These limiting measures will depend on which region of the disorder variables we are restricting ourselves to.\\
Let us look at the $k$-marginal
\begin{equation}\label{marginal}
\begin{split}
 \mu_{F,n}[\eta](\sigma_1,\dots,\sigma_k)&=\sum_{\o_{k+1},\dots,\o_n}\mu_{F,n}[\eta](\sigma_1,\dots,\sigma_k,\o_{k+1},\dots,\o_n)\cr
&=\sum_{\o_{k+1},\dots,\o_n}\frac{e^{-nF(L_n^{\sigma_{[1,k]},\o_{[k+1,n]}})}\prod_{i=1}^k\alpha[\eta_i](\sigma_i)\prod_{j=k+1}^n\alpha[\eta_j](\o_j)}{\sum_{\bar\sigma\in E^n}e^{-nF(L_n^{\bar\sigma})}\prod_{i=1}^n\alpha[\eta_i](\bar\sigma_i)}\cr
\end{split}
 \end{equation}
Let us now introduce the suitable decomposition of the empirical distribution, obtained by dividing the volume $\{1,\dots,n\}$ in two subvolumes $\{1,\dots,k\}$ and $\{k+1,\dots,n\}$, where $k$ is the size of the marginal we are considering; then we  focus on the respective $b$-like sites for both of the subvolumes.
\begin{equation}
\begin{split}\label{empiricalpart}
 L_n^{\sigma_{[1,k]},\o_{[k+1,n]}}&=\frac{k}{n}\frac{1}{k}\sum_{i=1}^k\d_{\sigma_i}+\frac{n-k}{n}\frac{1}{n-k}\sum_{i=k+1}^n\d_{\o_i}\cr
&=\frac{k}{n}\sum_{b\in E'}\hat\pi_{[1,k]}(b)\hat L_{[1,k]}(b)+\frac{n-k}{n}\sum_{b\in E'}\hat\pi_{[k+1,n]}(b)\hat L_{[k+1,n]}(b)
\end{split}
\end{equation}
In the process to carry out \eqref{empiricalpart}, we have also made use of the following definitions
\begin{equation}
 \begin{split}\label{partvolum}
&\Lambda_{[1,k]}(b)=\left\lbrace i\in \{1,\dots,k\}:\eta(i)=b\right\rbrace \cr  
&\Lambda_{[k+1,n]}(b)=\left\lbrace i\in \{k+1,\dots,n\}:\eta(i)=b\right\rbrace \cr
&\hat\pi_{[1,k]}(b)=\frac{|\Lambda_{[1,k]}(b)|}{k},\;\;\;\;\hat\pi_{[k+1,n]}(b)=\frac{|\Lambda_{[k+1,n]}(b)|}{n-k}
\end{split}
\end{equation}
{\bf Proof of Theorem \ref{thm1}.}
Let us start providing the key result, namely the {\em weak convergence} of the measure $\mu_{F,n}[\eta]$ to $\mu_j[\eta]=\prod_{i=1}^{\infty} \gamma[\eta(i)](\cdot |\pi\hat \nu_j)$ conditional on the suitable region of 
the disorder.\\
The following Lemma is the short view (local topology) version of Lemma \eqref{conti}.
\begin{lem}\label{conti_state}
 For any event $A$ which depends only on the first $k$ coordinates the following holds
\begin{equation}
 \mid \mu_{F,n}[\eta](A)-\mu_j[\eta](A)\mid\leq\tilde r(n)\;\;\;\;\forall \; \eta\in \HH_{j,n,k}^{\t}
\end{equation}
where 
$\lim_{n\uparrow\infty}\tilde r(n)=0$. 
\end{lem}
\bigskip

\noindent
\textbf{Proof: } It suffices to consider the event $A$ which fixes the first $k$ coordinates and write 
\begin{equation}
\begin{split}
 &\mu_{F,n}[\eta](\sigma_1,\dots,\sigma_k)=\cr
&=\sum_{\o_{k+1},\dots,\o_n}\frac{e^{-nF\left( \frac{k}{n}\displaystyle\sum_{b\in E'}\hat\pi_{[1,k]}(b)\hat L_{[1,k]}^{\sigma}(b)+\frac{n-k}{n}\displaystyle\sum_{b\in E'}\hat\pi_{[k+1,n]}(b)\hat L_{[k+1,n]}^{\o}(b)\right) }\displaystyle\prod_{i=1}^k\alpha[\eta_i](\sigma_i)\prod_{j=k+1}^n\alpha[\eta_j](\o_j)}{\displaystyle\sum_{\bar\sigma\in E^n}e^{-nF(L_n^{\bar\sigma})}\prod_{i=1}^n\alpha[\eta_i](\bar\sigma_i)}\cr
&=\tiny\frac{\displaystyle\sum_{\hat\nu\in \hat M_{[k+1,n]}}\sum_{{\o_{k+1},\dots,\o_n}:\atop{\hat L_{[k+1,n]}^{\tilde\o}(\cdot)=\hat\nu(\cdot)}}e^{-nF\left( \frac{k}{n}\left\langle\hat\pi_{[1,k]},\hat L_{[1,k]}^{\sigma}\right\rangle +\tiny\frac{n-k}{n}\left\langle\hat\pi_{[k+1,n]},\hat\nu\right\rangle \right) }\prod_{i=1}^k\alpha[\eta_i](\sigma_i)\prod_{j=k+1}^n\alpha[\eta_j](\o_j)}{\displaystyle\sum_{\bar\nu\in \hat M_n}\sum_{{\bar\sigma\in E^n:}\atop{\hat L_n^{\bar\sigma}=\bar\nu}}e^{-nF(L_n^{\bar\sigma})}\prod_{i=1}^n\alpha[\eta_i](\bar\sigma_i)}
\end{split}
\end{equation}
 where we have introduced the following space
\begin{equation}\label{space}
\hat M_{[k+1,n]}:=\left\lbrace \hat\nu\in \PP(E)^{E'}:\exists\; \tilde\o=(\tilde\o_{k+1},\dots,\tilde\o_n):\hat L_{[k+1,n]}^{\tilde\o}(b)=\hat\nu(b),\forall\;b\in E'\right\rbrace  
\end{equation}
Using the partition induced by the disorder variables $\eta$ on the sub-volume $\{k+1,\dots,n\}$, we have
\begin{equation}
\begin{split}
& \sum_{{\o_{k+1},\dots,\o_n}:\atop{\hat L_{[k+1,n]}^{\o}(\cdot)=\hat\nu(\cdot)}}\prod_{j=k+1}^n\alpha[\eta_j](\o_j)
=\prod_{b\in E'}\alpha[b]^{|\Lambda_{[k+1,n]}(b)|}(\O_{|\Lambda_{[k+1,n]}(b)|}(\hat\nu(b)))
\end{split}
\end{equation}
and with 
this we obtain
 \begin{equation}
\begin{split}
 &\mu_{F,n}[\eta](\sigma_1,\dots,\sigma_k)=\cr
&=\frac{\displaystyle\prod_{i=1}^k\alpha[\eta_i](\sigma_i)\sum_{\hat\nu\in \hat M_{[k+1,n]}}e^{-nF\left( \frac{k}{n}\left\langle\hat\pi_{[1,k]},\hat L_{[1,k]}^{\sigma}\right\rangle +\frac{n-k}{n}\left\langle\hat\pi_{[k+1,n]},\hat\nu\right\rangle \right) }\prod_{b\in E'}\alpha[b]^{|\Lambda_{[k+1,n]}(b)|}(\O_{|\Lambda_{[k+1,n]}(b)|}(\hat\nu(b)))}{\displaystyle\sum_{\nu'\in \hat M_{[1,k]}}\sum_{\hat\nu\in \hat M_{[k+1,n]}}\sum_{{\bar\sigma\in E^k:}\atop{\hat L_{[1,k]}^{\bar\sigma}=\nu'}}e^{-nF\left( \frac{k}{n}\left\langle\hat\pi_{[1,k]},\hat L_{[1,k]}^{\bar\sigma}\right\rangle +\frac{n-k}{n}\left\langle\hat\pi_{[k+1,n]},\hat\nu\right\rangle \right) }\prod_{i=1}^k\alpha[\eta_i](\bar\sigma_i)\sum_{{\tilde\sigma\in E^{n-k}:}\atop{\hat L_{[k+1,n]}^{\tilde\sigma}=\nu'}}\prod_{j=k+1}^n\alpha[\eta_j](\tilde\sigma_j)}
\end{split}
\end{equation}
 where naturally $\displaystyle\prod_{b\in E'}\alpha[b]^{|\Lambda_{[k+1,n]}(b)|}(\O_{|\Lambda_{[k+1,n]}(b)|}(\hat\nu(b)))=\sum_{{\tilde\sigma\in E^{n-k}:}\atop{\hat L_{[k+1,n]}^{\tilde\sigma}=\nu'}}\prod_{j=k+1}^n\alpha[\eta_j](\tilde\sigma_j)$.

Now multipling and dividing, both numerator and denominator, by $e^{-nF(\left\langle\hat\pi_{[k+1,n]},\hat\nu\right\rangle)}$ we arrive at:
\begin{equation}\label{diffformMU}
\begin{split}
&\mu_{F,n}[\eta](\sigma_1,\dots,\sigma_k)=\cr
&=\frac{\displaystyle\prod_{i=1}^k\alpha[\eta_i](\sigma_i)\sum_{\hat\nu\in \hat M_{[k+1,n]}}e^{-n\left[ F\left( \frac{k}{n}\left\langle\hat\pi_{[1,k]},\hat L_{[1,k]}^{\sigma}\right\rangle +\frac{n-k}{n}\left\langle\hat\pi_{[k+1,n]},\hat\nu\right\rangle \right) -F(\left\langle\hat\pi_{[k+1,n]},\hat\nu\right\rangle)\right] }\rho_{F,n,k}[\eta](\hat\nu)}{\displaystyle\sum_{\nu'\in \hat M_{[1,k]}}\sum_{\hat\nu\in \hat M_{[k+1,n]}}\sum_{{\bar\sigma\in E^k:}\atop{\hat L_{[1,k]}^{\bar\sigma}=\nu'}}e^{-n\left[ F\left( \frac{k}{n}\left\langle\hat\pi_{[1,k]},\hat L_{[1,k]}^{\bar\sigma}\right\rangle +\frac{n-k}{n}\left\langle\hat\pi_{[k+1,n]},\hat\nu\right\rangle\right) -F(\left\langle\hat\pi_{[k+1,n]},\hat\nu\right\rangle)\right]}\prod_{i=1}^k\alpha[\eta_i](\bar\sigma_i)\rho_{F,n,k}[\eta](\hat\nu)}
\end{split} 
\end{equation}
 where $\rho_{F,n,k}[\eta]\in \PP(\hat M_{[k+1,n]})$ is defined as
\begin{equation}\label{ro}
 \rho_{F,n,k}[\eta](\hat\nu):=\frac{e^{-nF(\left\langle\hat\pi_{[k+1,n]},\hat\nu\right\rangle)}\prod_{b\in E'}\alpha[b]^{|\Lambda_{[k+1,n]}(b)|}(\O_{|\Lambda_{[k+1,n]}(b)|}(\hat\nu(b)))}{\sum_{\tilde\nu\in \hat M_{[k+1,n]}}e^{-nF(\left\langle\hat\pi_{[k+1,n]},\hat\nu\right\rangle)}\prod_{b\in E'}\alpha[b]^{|\Lambda_{[k+1,n]}(b)|}(\O_{|\Lambda_{[k+1,n]}(b)|}(\tilde\nu(b)))}
\end{equation}

\vskip .1in
\noindent
Note that the measure $\rho_{F,n,k}[\eta]$ depends on the disorder variables just in the subvolume $\{k+1,\dots,n\}$.
\bigskip


Recall that a 
function $F:\PP(E)\rightarrow \R$ is differentiable if, for all $\alpha \in \PP(E)$ there is a linear map $dF_{\alpha}:T(\PP(E))\rightarrow \R$ on the tangent space such that
\begin{equation}
 F(\alpha')=F(\alpha)+dF_{\alpha}(\alpha'-\alpha)+||\alpha'-\alpha|| r(\alpha',\alpha)
\end{equation}
where $\alpha'\rightarrow r(\alpha',\alpha)$ is continuous at $\alpha'=\alpha$ with $r(\alpha,\alpha)=0$.
Then uniformly in $\a,\a'$ we have 
\begin{equation}
 \sup_{\a,\a'}|F(\a+p(\a'-\a))-F(\a)-p dF_{\a}(\a'-\a)|\leq Cpr(p)
\end{equation}
where $r(p)\downarrow 0$ with $p\downarrow 0$. The uniformity in $\a,\a'$ follows by the compactness of $\PP(E)$.
In our set up we will have
\begin{equation}
 |F\left( \frac{k}{n}L_{[1,k]}(\sigma)+\frac{n-k}{n}\nu\right) -F(\nu)-dF_{\nu}\frac{k}{n}(L_{[1,k]}(\sigma)-\nu)|\leq C\frac{k}{n}r(\frac{k}{n})
\end{equation}
where we have set $\nu=\left\langle\hat\pi_{[k+1,n]},\hat\nu\right\rangle$. 
This gives, recognizing that $\left\langle\hat\pi_{[1,k]},\hat L_{[1,k]}^{\sigma}\right\rangle$ and $\nu$ are both elements in $\PP(E)$, the upper bound
\begin{equation}
 \begin{split}\label{bound11}
&\mu_{F,n}[\eta](\sigma_1,\dots,\sigma_k)\cr
&\leq\frac{e^{2Ckr(\frac{k}{n})}\displaystyle\prod_{i=1}^k\alpha[\eta_i](\sigma_i)\sum_{\hat\nu\in \hat M_{[k+1,n]}}\prod_{i=1}^k e^{-dF_{\left\langle\hat\pi_{[k+1,n]},\hat\nu\right\rangle}(\d_{\sigma_i}-\left\langle\hat\pi_{[k+1,n]},\hat\nu\right\rangle)}\rho_{F,n,k}[\eta](\hat\nu)}{\displaystyle\sum_{\hat\nu\in \hat M_{[k+1,n]}}\rho_{F,n,k}[\eta](\hat\nu)\prod_{i=1}^k\sum_{\bar\sigma_i\in E}\alpha[\eta_i](\bar\sigma_i) e^{-dF_{\left\langle\hat\pi_{[k+1,n]},\hat\nu\right\rangle}(\d_{\bar\sigma_i}-\left\langle\hat\pi_{[k+1,n]},\hat\nu\right\rangle)}}
  \end{split}
\end{equation}

and the corresponding lower bound which is obtained from the last r.h.s. by replacing $C>0$ by $-C$. 
The measure $\rho_{F,n,k}[\eta]$ can be written in the form
\begin{equation}
 \rho_{F,n,k}[\eta](\hat\nu)=\frac{\rho_{F,n-k}[\eta](\hat\nu)e^{-kF(\left\langle\hat\pi_{[k+1,n]},\hat\nu\right\rangle)}}{\displaystyle\sum_{\bar\nu\in \hat M_{[k+1,n]}}\rho_{F,n-k}[\eta](\bar\nu)e^{-kF(\left\langle\hat\pi_{[k+1,n]},\bar\nu\right\rangle)}}
\end{equation}
where we have recovered the proper random mean-field measure on the empirical distribution of size $n-k$. 
%
Note once again that $\left\langle\hat\pi_{[k+1,n]},\hat\nu\right\rangle \in \PP(E)$, and when $\hat\nu$ moves in $\hat M_{[k+1,n]}$ the corresponding measure $\left\langle\hat\pi_{[k+1,n]},\hat\nu\right\rangle$ is moving among the possible empirical measures of size $n-k$.
Let us write  $C(A,\e):=\{\nu\in\PP(E)^{E'}:d(\nu,A)\leq\e\}$, for the $\e$-ball of a set $A$. By definition $\rho_{F,n}[\eta]$ is said to concentrate on the set $A$ iff $\rho_{F,n}[\eta](C(A,\e)^c)\downarrow0,\;\forall\;\e>0$. So whenever $\rho_{F,n}[\eta]$ concentrates on a finite set, so does $\rho_{F,n,k}[\eta]$, by the boundedness of $F$.\\
We remark that $\rho_{F,n-k}[\eta](\hat\nu)=\mu_{F,n-k}[\eta](L_{[k+1,n]}=\left\langle\hat\pi_{[k+1,n]},\hat\nu\right\rangle)$ 
has the property of concentrating around the minimizers, as we know 
from the analysis of the previous chapter.  \bigskip

To study the bound \eqref{bound11} and the corresponding lower bound let's introduce the quantities
\begin{equation}\label{csi}
\begin{split} 
&\xi_N=\displaystyle\sum_{\hat\nu\in \hat M_{[k+1,n]}}\rho_{F,n,k}[\eta](\hat\nu)\prod_{i=1}^k\alpha[\eta_i](\sigma_i)\prod_{i=1}^k e^{-dF_{\left\langle\hat\pi_{[k+1,n]},\hat\nu\right\rangle}(\d_{\sigma_i}-\left\langle\hat\pi_{[k+1,n]},\hat\nu\right\rangle)}\cr
&\xi_D=\displaystyle\sum_{\hat\nu\in \hat M_{[k+1,n]}}\rho_{F,n,k}[\eta](\hat\nu)\prod_{i=1}^k\sum_{\tilde\sigma_i\in E}\alpha[\eta_i](\tilde\sigma_i)\prod_{i=1}^k e^{-dF_{\left\langle\hat\pi_{[k+1,n]},\hat\nu\right\rangle}(\d_{\tilde\sigma_i}-\left\langle\hat\pi_{[k+1,n]},\hat\nu\right\rangle)}
\end{split}
\end{equation}
 Let us decompose the $\sum_{\hat\nu\in \hat M_{[k+1,n]}}$ over $C(M^*,\e)$ and its complement, and compare the terms with their values at the midpoints:
\begin{equation}
 \begin{split}\label{decomposition}
  \xi_N&=\displaystyle\sum_{\hat\nu^*\in M^*}\sum_{{\hat\nu\in}\atop {B(\e,\hat\nu^*)\cap \hat M_{[k+1,n]}}}\rho_{F,n,k}[\eta](\hat\nu)\prod_{i=1}^k\alpha[\eta_i](\sigma_i)\prod_{i=1}^k e^{-dF_{\pi\hat\nu^*}(\d_{\sigma_i}-\pi\hat\nu^*)}+\cr
&+\displaystyle\sum_{\hat\nu^*\in M^*}\sum_{{\hat\nu\in}\atop {B(\e,\hat\nu^*)\cap\hat M_{k+1,n}}}\rho_{F,n,k}[\eta](\hat\nu)\prod_{i=1}^k\alpha[\eta_i](\sigma_i)\left[ \prod_{i=1}^k 
\tiny e^{-dF_{\left\langle\hat\pi_{[k+1,n]},\hat\nu\right\rangle}(\d_{\sigma_i}-\left\langle\hat\pi_{[k+1,n]},\hat\nu\right\rangle)}-\prod_{i=1}^k
\tiny e^{-dF_{\pi\hat\nu^*}(\d_{\sigma_i}-\pi\hat\nu^*)}\right] +\cr
&+\displaystyle\sum_{{\hat\nu^\in}\atop {C(M^*,\e)^c\cap\hat M_{[k+1,n]}}}\rho_{F,n,k}[\eta](\hat\nu)\prod_{i=1}^k\alpha[\eta_i](\sigma_i)\prod_{i=1}^k e^{-dF_{\left\langle\hat\pi_{[k+1,n]},\hat\nu\right\rangle}(\d_{\sigma_i}-\left\langle\hat\pi_{[k+1,n]},\hat\nu\right\rangle)}\cr
 \end{split}
\end{equation}
The sum in the last line is bounded by a function $r_{\text{cp}}(\e,n)$, 
where $\lim_{n\uparrow\infty}r_{\text{cp}}(\e,n)=0$, when $\eta$ is in the union of the {\em good-sets}.
This holds by the concentration property of the empirical distribution given in  
\eqref{concentrantion-prop} applied to the measure for sites $\geq k$, 
using the boundedness of the first derivative of $F$. 

The second line is bounded in modulus by a function $\g(\e)$ where 
$\lim_{\e\downarrow 0}\g(\e)=0$ by the twice continuous differentiability of $F$.

This implies the bounds 

\begin{equation}
\begin{split}
 &\Bigl | \xi_N - \sum_{\hat\nu^*\in M^*}\rho_{F,n,k}[\eta](\tilde{B}(\e,\hat\nu^*))\prod_{i=1}^k 
 \alpha[\eta_i](\sigma_i)
  e^{-dF_{\pi\hat\nu^*}(\d_{\sigma_i}-\pi\hat\nu^*)}\Bigl | \leq \g(\e)+r_{cp}(\e,n)\cr
\end{split}
\end{equation}
assuming that $\eta$ is in the union of the {\em good-sets}. 

Summing over the finitely many values of $\s_1, \dots,\s_k$ we obtain the same 
type of bounds (with possibly worse functions $\g(\e),r_{cp}(\e,n)$) for $\x_D$. 

Recall the definition of the kernels \eqref{colonel} and choose the disorder variable
$\eta\in \HH^\t_{i,n,k}$ in the part 
of the {\em good-set} which ensures the dominance of the $i$-th minimizer. This gives 
that 
\begin{equation}\label{absvalue}
 \mid\mu_{F,n}[\eta](\sigma_1,\dots,\sigma_k)-\displaystyle\prod_{j=1}^k\gamma[\eta_j](\sigma_j|\pi\hat\nu_i^*) \mid\leq\; \zeta(\e)+\;\chi(\e,n)
\end{equation}
where $\lim_{n\uparrow\infty}\chi(\e,n)=0$ and $\lim_{\e\downarrow 0}\zeta(\e)=0$.
This proves the lemma and also the statement \eqref{uniform}.
$\Cox$
\bigskip

\begin{lem}\label{limH_n(L)_state}
Let $\Xi$ be a continuous real-valued function on $\PP(E^\infty)\times (E')^{\infty}$. Then under the {\em non-degeneracy assumptions} 1) and 2) the following holds:
\begin{equation}
\lim_{n\uparrow\infty}\int_{\HH_{i,n}^{\delta_n}}\P_\pi(d\eta)\Xi(\mu_{F,n}[\eta],\eta)=w_i\int \P_\pi(d\eta)(d\eta)\Xi(\mu_i[\eta],\eta)
\end{equation}
$w_i=\P_{\pi}(G\in R_i)$.
\end{lem}
The proof of this lemma, thanks to the continuity of $\Xi$ which allows finite dimensional approximation 
and to the Lemma\eqref{conti_state}, follows the trail drawn by the proof of the Lemma\eqref{limH_n(L)}.

Using all the tools we have provided, we find
\begin{equation}
 \lim_{n\uparrow\infty}\int\P_{\pi}(\eta)\Xi(\mu_{F,n}[\eta],\eta)=\sum_{j=1}^k\int\P_{\pi}(\eta)\Xi(\mu,\eta)w_j\d_{\mu_j[\eta]}(d\mu)
\end{equation}
where we can identify
\begin{equation}
\begin{split}
 & J(d\mu,d\eta)=\sum_{j=1}^k\P_{\pi}(d\eta)w_j \d_{\mu_j[\eta]}(d\mu)\cr
&\Longrightarrow J(d\mu|\eta)=\sum_{j=1}^kw_j\d_{\mu_j[\eta]}(d\mu)
\end{split}
\end{equation}
This finishes the proof of the Main Theorem \ref{thm1} and concludes the paper.  
$\Cox$
\bigskip

\subsection{Acknowledgements}
We thank Aernout van Enter, Marco Formentin, Alex Opoku, and Victor Ermolaev 
for stimulating discussions and the NWO for support under project number 613.000.606.

\end{document}